\documentclass[reprint, 
superscriptaddress,
 amsmath,amssymb,
 aps, 
pra,
floatfix
]{revtex4-2}

\usepackage{xfrac}
\usepackage{lmodern}
\usepackage{graphicx}
\usepackage{dcolumn}
\usepackage{bm}
\usepackage{hyperref}

\hypersetup{
	colorlinks = true,
	urlcolor   = blue,
	linkcolor  = blue,
	citecolor  = blue
}

\DeclareMathOperator{\LG}{LG}
\DeclareMathOperator{\erf}{erf}
\DeclareMathOperator{\erfc}{erfc}
\DeclareMathOperator{\sinc}{sinc}
\newcommand*{\bra}[1]{\langle #1 |} 
\newcommand*{\ket}[1]{| #1 \rangle} 

\long\def\comment#1{}

\usepackage[T1]{fontenc}
\usepackage[utf8x]{inputenc}
\usepackage{graphicx}
\usepackage[dvipsnames]{xcolor}
\usepackage{upgreek}

\begin{document}

\preprint{APS/123-QED}

\title{\textbf{Addressable Rydberg excitation in arrays of single neutral atoms with a strongly focused flat-top beam} 
}

\author{I. V. Iukhnovets}
\email{Contact author: i.yukhnovets@rqc.ru}
\affiliation{Moscow Institute of Physics and Technology (MIPT), 141700 Dolgoprudny, Moscow Region, Russia}
\affiliation{Russian Quantum Center (RQC), 143025 Skolkovo, Russia}
\affiliation{P. N. Lebedev Physical Institute (LPI), 119991 Moscow, Russia}
\author{M. Y. Goloshchapov}
\affiliation{Technical University of Munich, Arcisstraße 21, 80333, Munich, Germany}
\affiliation{Ludwig-Maximilians-Universität München, Geschwister-Scholl-Platz 1, 80539, Munich, Germany}
\author{A. P. Gordeev} 
\affiliation{P. N. Lebedev Physical Institute (LPI), 119991 Moscow, Russia}
\affiliation{Quantum Technology Centre and Faculty of Physics, M. V. Lomonosov Moscow State University, 119991 Moscow,
Russia}
\author{O.~V.~Bychkova}%
\affiliation{P. N. Lebedev Physical Institute (LPI), 119991 Moscow, Russia}
\author{I. B. Bobrov}%
\affiliation{Quantum Technology Centre and Faculty of Physics, M. V. Lomonosov Moscow State University, 119991 Moscow,
Russia}
\author{G. I. Struchalin}%
\affiliation{Quantum Technology Centre and Faculty of Physics, M. V. Lomonosov Moscow State University, 119991 Moscow,
Russia}
\author{S. S. Straupe}%
\affiliation{Russian Quantum Center (RQC), 143025 Skolkovo, Russia}
\affiliation{Quantum Technology Centre and Faculty of Physics, M. V. Lomonosov Moscow State University, 119991 Moscow,
Russia}

\date{\today}

\begin{abstract}

We present a method for generating a laser beam with flat intensity and phase profiles in the focal region where the beam interacts with neutral $^{87}$Rb atoms in an array of optical dipole traps. We synthesize the beam as a superposition of Hermite–Gaussian or Laguerre–Gaussian modes. Then we give analytical expressions for the coefficients of such a superposition, an analysis of beam propagation along the $z$ axis in the vicinity of the waist, and several other related theoretical issues. Rydberg two-qubit dynamics driven by this flat-top profile are analyzed through numerical solutions of the Lindblad master equation using our in-house Julia package. Beam preparation is demonstrated on a neutral-atom experimental platform. Measurements reveal a difference in the visibility of Rabi oscillations for addressed atoms compared with neighboring ones, confirming the effective spatial selectivity provided by the flat-top beam profile.
\end{abstract}

\maketitle


\section{Introduction \label{sec:intro}}

The development of quantum computers based on neutral atoms is currently an active area of research due to the possibility of scaling quantum registers to several thousand qubits \cite{Pause:2023pao,Manetsch:2024lwl,Chiu:2025uis}. A major focus in this field is the realization of scalable, high-fidelity single- and multi-qubit entangling gates. One important class of atomic quantum computing platforms is based on Rydberg excitation, in which a highly excited Rydberg state, typically denoted $\ket{r}$, is used to implement quantum gates. A key advantage of such systems is the Rydberg blockade effect, whereby excitation of one atom to $\ket{r}$ shifts the Rydberg levels of nearby atoms and thereby suppresses their simultaneous excitation. This mechanism provides the physical basis for the implementation of entangling gates. The transition to $\ket{r}$ is driven via a two-photon process with detuning $\Delta$ from an intermediate state $\ket{p}$, which serves to suppress population of the intermediate state. \par
Improving the fidelity of Rydberg two-qubit operations remains an important and ongoing challenge. One of the factors limiting gate quality is the uniformity of the beam intensity and phase profile across the cross section of the dipole traps containing the target atoms. \par
In Ref.\,\cite{Levine:2019zfq}, entanglement was achieved using global Rydberg beams applied to atoms within a dedicated entanglement zone. Ten atoms were previously moved there from a storage zone. In this way, five entangled pairs were demonstrated, and CZ gates were implemented in parallel. A similar separation between storage and entanglement regions is employed in Refs. \cite{Bluvstein:2023zmt,Chiu:2025uis,Evered:2025zza}. \par
\begin{figure}[t]
\includegraphics[height=0.3\textwidth]{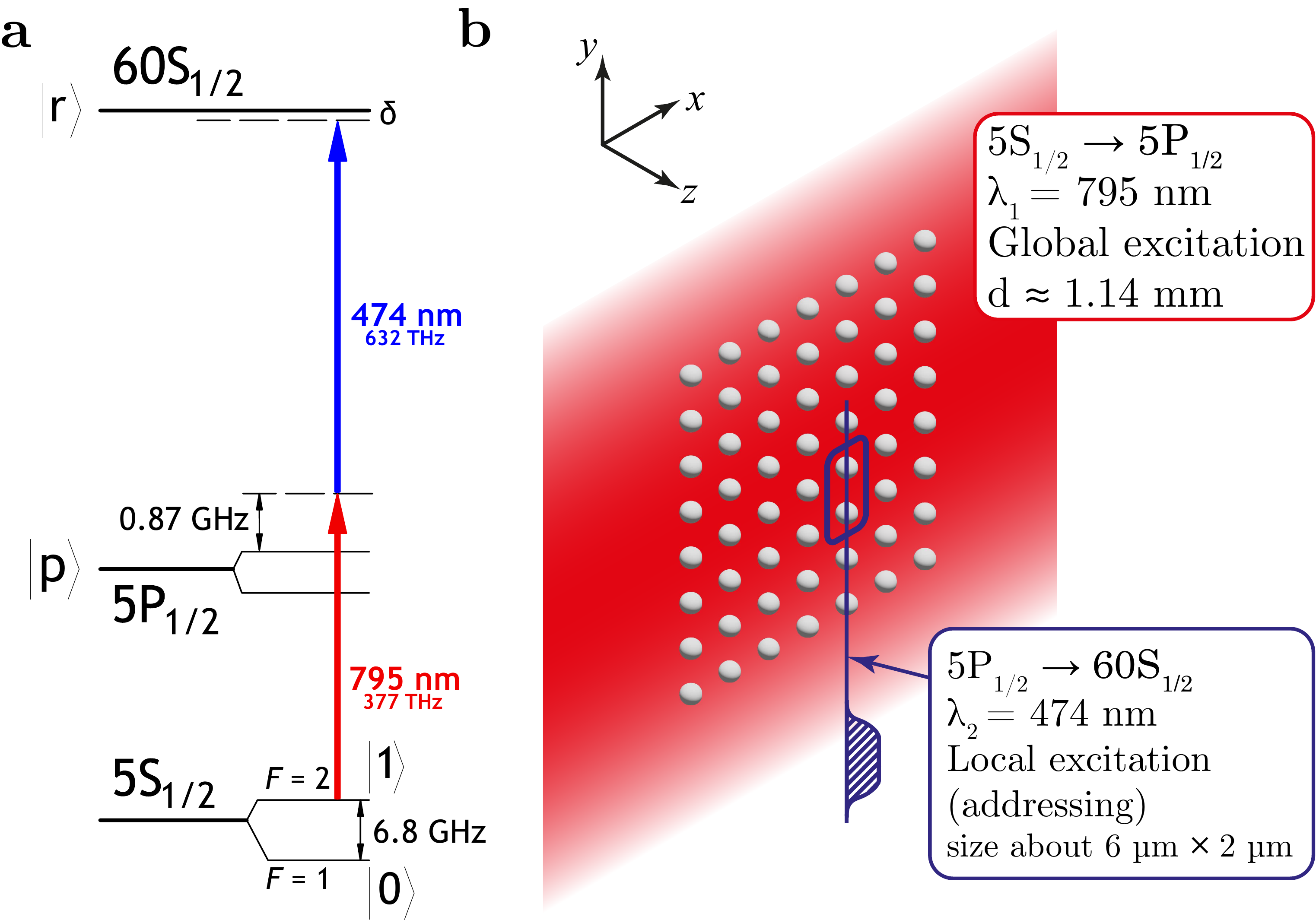}
\caption{\label{Flattop} Illumination of atoms in a trap array by two Rydberg-excitation beams. (a) Scheme of two-photon excitation to the $\ket{r}$ state via the intermediate $\ket{p}$ state. (b) Experimental realization. The beam for the first excitation stage is global and propagates along the $OX$ direction. The  for the second excitation stage provides site-selective addressing, propagates along $OZ$, and has a flat-top intensity profile.}
\end{figure}

In contrast, the present work employs local addressing. The main advantage of this approach is the reduced gate time, since atoms do not need to be transported between zones. This can increase the achievable circuit depth within the coherence time of the system. In our implementation with $^{87}$Rb, the first stage of the Rydberg excitation is provided by a global beam illuminating the entire qubit array, while the second stage is driven by a tightly focused addressing beam. The intermediate state is chosen to be $5P_{1/2}$ and the Rydberg state is $60S_{1/2}$ (Fig.~\ref{Flattop}a). The addressing beam covers only the two target atoms while minimizing crosstalk to neighboring atoms~(Fig.~\ref{Flattop}b).

Addressed Rydberg excitation using Gaussian beams is a well-established technique. Approaches can be broadly divided into two categories: using a beam waist large enough to illuminate both atoms with equal intensity \cite{Ma:2023ltx,Fu:2021ybt,Ocola:2022xdd}, or exciting each atom with an individual strongly focused beam \cite{Graham:2019cfg,Radnaev:2024lyk,Bornet:2024sjy,Li:2024qtg}. In the former case, the beam waist is typically on the order of 10 $\upmu$m, whereas in the latter it is below 5 $\upmu$m. A clear drawback of the former approach is that the addressing light also illuminates neighboring atoms, leading to crosstalk (undesired changes in their quantum states). \par
Another approach for improving the fidelity of quantum operations at nonzero atomic motional temperature in an optical dipole trap is the use of laser beams with a flat intensity and phase profile. For brevity, we refer to such a beam as a flat-top beam. The beam intensity and phase are required to remain uniform over the spatial region in which the atom is localized within the trap. Outside this region, the detailed structure of the electromagnetic field is less critical, but it is sufficient for the field amplitude to decay rapidly toward the periphery to suppress crosstalk with neighboring atoms. \par
Several studies have reported the use of dedicated beam-shaping optical elements to generate beams with a flat-top intensity distribution across most of the illuminated area. For example, Ref.\,\cite{Mielec:2018fsj} describes a long-interrogation-time cold-atom interferometer, in which an Asphericon TSM-25-10-S-B beam shaper is employed. The resulting beam exhibits an approximately uniform intensity over a 28 mm wide region with rms variations of about 10\%. In Ref.\,\cite{Pause2023}, an AdlOptica Focal-$\pi$Shaper Q is used to illuminate a microlens array forming dipole traps. \par
In Ref.\,\cite{Ebadi:2020ldi}, a beam with a flat-top intensity profile is generated using a spatial light modulator (SLM). The authors employ a conjugate-gradient method and iteratively refine the hologram to improve the profile uniformity. They note that a trade-off must be found between profile uniformity and optical power loss. The achieved conversion efficiency does not exceed 40\%. \par
During the earlier stages of this research, we used a flat-top beam shaper in our setup for addressed atomic excitation. The method presented below allowed us to achieve a higher degree of uniformity in the flat region of the beam than was possible with that technique. In the present work, we develop an improved analytical description of a flat-top profile and employ a spatial light modulator to generate a flat-top beam for addressed Rydberg excitation. Unlike Ref.\,\cite{Ebadi:2020ldi}, we construct the SLM hologram for the flat-top profile obtained with our method using the algorithm proposed in Ref.\,\cite{Davis:1999sfk} and later refined in Ref.\,\cite{Bolduc:2013icd}. \par
The paper is organized as follows: in Secs.~\ref{sec:theory}--\ref{sec:hologram} we outline the theory behind our flat-top beam generation method and hologram computing; next, in Sec.~\ref{sec:model} we describe the model developed for calculating the parameters of Rydberg gates; in Sec.~\ref{sec:setup} we present the experimental setup; we then present the results; and finally, in Sec.~\ref{sec:conclusion}, we summarize our conclusions. Appendixes~\ref{sec:app:AnalyticalFlatTop} and \ref{sec:app:TaylorExpansion} discuss several special aspects of the theory behind flat-top beam preparation, while Appendix~\ref{sec:app:params} is devoted to the measurement of the parameters required for simulations carried out with our numerical package.

\section{Theory of the flat-top beam formation \label{sec:theory}}
A phase-only SLM allows one to imprint an arbitrary phase $\varphi(x,y)$ onto the wavefront of an incident beam, where $x$ and $y$ denote transverse coordinates in the plane perpendicular to the propagation direction. An input field $E_{\mathrm{in}}(x,y)$ after reflection from the liquid-crystal surface of the modulator acquires the phase phactor: $E_{\mathrm{out}}(x,y)=E_{\mathrm{in}}(x,y)\exp \left[i\varphi(x,y)\right]$. In the far field (corresponding to the plane of the atomic traps), the resulting amplitude is given by
\begin{equation}
E_{\mathrm{out}}=\mathcal{F}\left[E_{\mathrm{in}}(x,y) \exp \left[ i \varphi(x,y) \right] \right],
\end{equation}
where $\mathcal{F}\left[ \dots \right]$ denotes the Fourier transform.
\begin{figure}[t]
\includegraphics[height=0.3\textwidth]{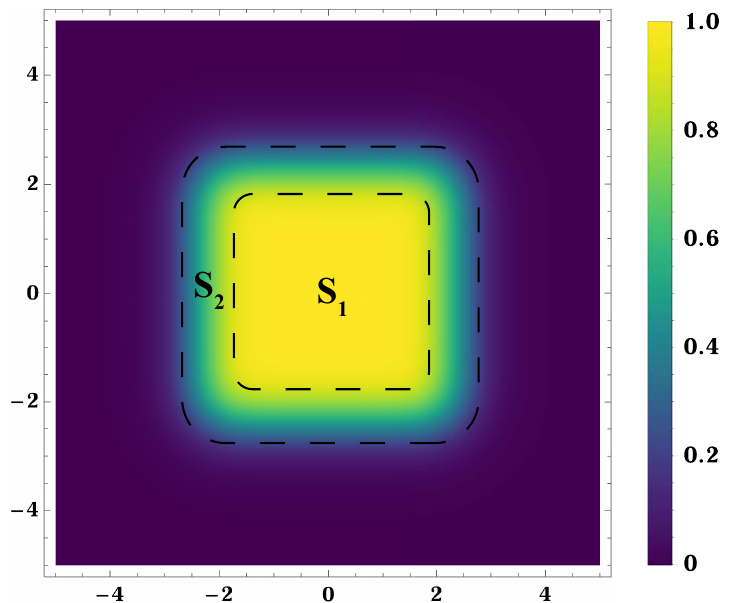}
\caption{\label{Zones} Flat-top profile zones (the color scale represents the intensity normalized to its maximum value)}
\end{figure}

Our goal is to determine a phase hologram to be displayed on the SLM that transforms an incident Gaussian beam $E_{\mathrm{in}}$ into a flat-top beam in the far field, while simultaneously incorporating a blazed diffraction grating so that the desired intensity distribution appears in the first diffraction order. This separates the target beam from parasitic radiation due to unmodulated reflection from the SLM surface. We note that an exact solution with 100\% efficiency (i.e., without power loss) does not exist, even theoretically, since the inverse Fourier transform of the desired flat-top beam cannot match the modulus of the initial Gaussian beam $E_{\mathrm{in}}$ \cite{Swan:2024ofh}. \par
We now specify the requirements for a beam to qualify as flat-top beam. Analytically, within the region $\mathbf{S_1}$ (Fig.~\ref{Zones}), both the intensity and the phase remain uniform. In practice, deviations from this ideal behavior are unavoidable and determine the beam quality. We refer to the region $\mathbf{S_2}$ as the “skirt”. Its geometric extent is constrained to suppress crosstalk with neighboring atoms.
\begin{figure*}[!htbp]
\includegraphics[width=0.96\textwidth]{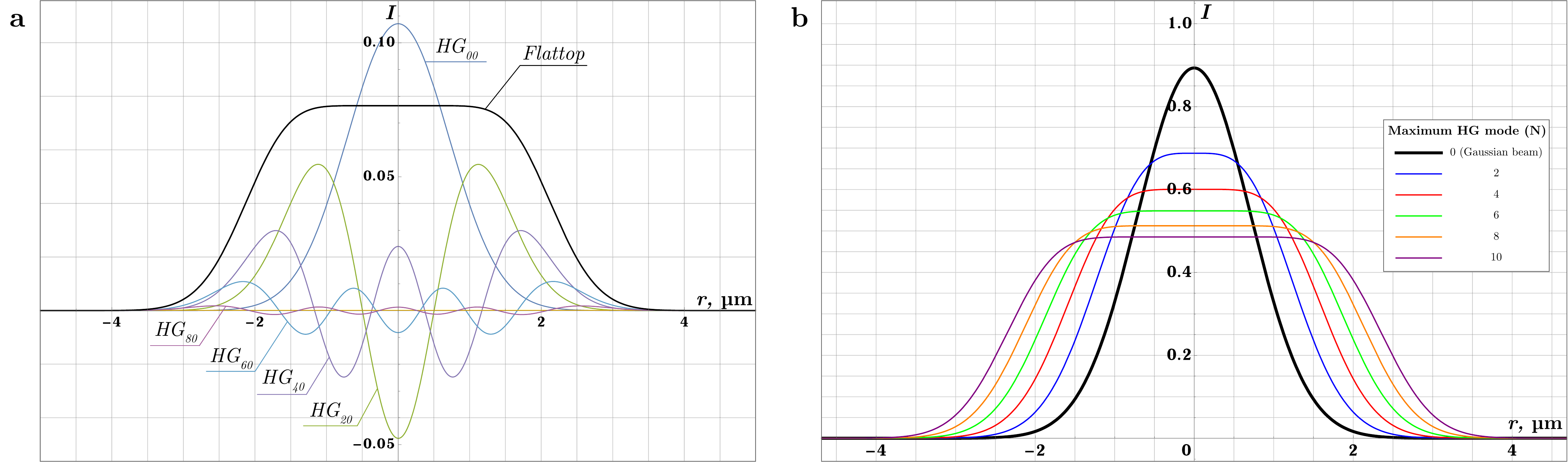}
\caption{\label{FlattopSuperp} (a) Flat-top beam constructed as a superposition of the first five even Hermite–Gaussian modes. (b) Dependence of the flat-top width and intensity on $N$.}
\end{figure*}

Various solution methods have been proposed in the literature. In \cite{Romero_JOSAA1995}, the phase profile of an aspheric lens $\varphi(x, y)$ that converts a Gaussian beam into a beam with a flat intensity profile without power loss is obtained analytically:
\begin{gather}
	\varphi(x, y) = c \varphi(x) \varphi(y),\notag\\ \quad \varphi(x) = \frac{\sqrt{\pi}}{2}  x \erf(x) + \frac{1}{2} \exp(-x^2) - \frac{1}{2}, \label{GeometricFlatTop}
\end{gather}
where $c$ denotes the lens curvature, which determines the transverse size of the flat beam, and $\erf(x)$ is the error function. Here and in what follows we assume that coordinates $x, y$ are normalized to some characteristic size. The advantages of this method include the absence of power loss, good intensity flatness, computational simplicity, and straightforward adjustment to the sizes of the input Gaussian beam and the output flat beam. A significant drawback, however, is the nonuniform phase profile, which to first order has a parabolic form. \par
An idealized choice for a flat beam would be $E(x, y) = E(x) E(y)$ with
\begin{equation}
	E(x) = \begin{cases}
		1, |x| \le 1/2, \\
		0, |x| > 1/2,
	\end{cases} \label{IdealFlatTop}
\end{equation}
where the full width at half maximum (FWHM) is unity by agreement. However, this choice is undesirable due to the presence of a sharp boundary at $|x| = 1/2$. If~(\ref{IdealFlatTop}) is used as the target field in iterative algorithms, the diffraction efficiency is typically quite low. The flat beam formed by the lens~(\ref{GeometricFlatTop}) likewise does not exhibit a sharp boundary. This behavior follows from fundamental diffraction constraints: producing a discontinuous edge would require optical elements and an incident Gaussian beam of infinite extent, which is infeasible. One is therefore led to the problem of finding a flat-beam profile with a smoothed boundary whose shape is physically well motivated. \par
Within the paraxial approximation, the wave equation allows separation of variables along the $x$ and $y$ axes; therefore, we restrict attention to fields of the form $E(x,y)=E(x)E(y)$. The beam profile is thus reduced to a one-dimensional function $E(x)$. The eigenfunctions of the paraxial wave equation in Cartesian coordinates are the Hermite–Gaussian modes $\mathrm{HG}_n$ of order $n$:
\begin{gather}
\mathrm{HG}_n(x)=H_n (\sqrt{2}x) \exp{(-x^2)}, \\
H_n(x)=(-1)^n \exp{(x^2)} \frac{\mathrm{d}^n \exp{(-x^2)}}{\mathrm{d}x^n},\ n \ge 0,
\end{gather}
where $H_n(x)$ denotes the Hermite polynomial of order $n$. \par
We seek the flat-top beam profile $E(x)$ in the form of a superposition of the lowest-order Hermite--Gaussian modes whose order does not exceed $N$:
\begin{equation}
E(x)=\sum_{n=0}^{N} \tilde c_n \mathrm{HG}_n (x),
\label{HGsum}
\end{equation}
where $\tilde c_n$ are real coefficients to be determined. This representation is preferable, for example, to a “super-Gaussian” distribution $\exp \left(-x^{2p} \right)$ \cite{Zhang:2016gil,Parent:1992jit,Jabczynski:208pol,Suresh:2025grf}, since a super-Gaussian is not physically well-motivated and is used for mathematical simplicity. \par
The flatness requirement for $E(x)$ is imposed by setting the lowest $K$ derivatives to zero at $x = 0$:
\begin{gather}
	\frac{d^k E(0)}{dx^k} = 0, \quad k = 1, \dots, K, \label{ZeroDeriv} \\
	E(0) = 1, \label{Normalization}
\end{gather}
where the final equation is included to fix the normalization of $E(x)$. Clearly, the system~(\ref{ZeroDeriv}--\ref{Normalization}) is linear in~$\tilde c_n$. The number of derivatives $K$ should be chosen as large as possible for a fixed $N$, provided that the system admits a solution. \par
Figure~\ref{FlattopSuperp}a shows the flat-top beam profile obtained for $N=8$, while Fig.~\ref{FlattopSuperp}b presents the dependence of the flat-top width and intensity on $N$. \par
We obtain an explicit solution of the system~(\ref{ZeroDeriv}--\ref{Normalization}) for arbitrary $N$ with $K = N$, thereby determining the flat-beam profile $E(x)$. A straightforward but algebraically involved solution is presented in Appendix \ref{sec:app:AnalyticalFlatTop}. Here, in the main text, we deduce the solution using Taylor series, although it is less rigorous. \par
Let us rewrite~(\ref{HGsum}), explicitly separating the polynomial part:
\begin{equation}
	E(x) = \exp (-x^2) \sum_{n = 0}^N \tilde c_n H_n(\sqrt{2} x). \label{SuperposHermite}
\end{equation}
We now examine in more detail the set of functions of the form~(\ref{SuperposHermite}) with arbitrary coefficients $\tilde c_n$. By the construction of the problem, the desired solution of the system~(\ref{ZeroDeriv}--\ref{Normalization}) belongs to this set. A superposition of Hermite polynomials $H_n$ is itself a polynomial of order $N$. Conversely, any polynomial of order $N$ can be expanded in the basis of $H_n$, since the $H_n$ form a complete orthogonal system on the real line. Therefore, instead of~(\ref{SuperposHermite}), the solution may be sought within the equivalent set of functions:
\begin{equation}
	E(x) = \exp (-x^2) \sum_{n = 0}^N c_n x^n, \label{SuperposPoly}
\end{equation}
where $c_n$ are new coefficients that depend linearly on $\tilde c_n$. We note that the factor $\sqrt{2}$ multiplying $x$ has been absorbed into $c_n$. \par
We represent (\ref{SuperposPoly}) as $E(x) = p_N(x) g(x)$, where $p_N(x)$ is an arbitrary polynomial of order $N$ and $g(x) \equiv \exp(-x^2)$. The functions $p_N(x) g(x)$ are entire and can therefore be expanded in a convergent Taylor series. From the system of equations~(\ref{ZeroDeriv}--\ref{Normalization}), it immediately follows that the solution $E(x)$ has the Taylor expansion:
\begin{equation}
	E(x) = 1 + o(x^K), \label{SolutionTaylor}
\end{equation}
where $o(x^K)$ denotes a term of lower order than $x^K$. \par
Consider the identity $\frac{1}{g(x)} \times g(x) = 1$. Expanding the first factor in a Taylor series gives $1/g(x) = \exp(x^2) = q_N(x) + o(x^N)$, where the “leading” polynomial $q_N(x) = \sum_{n=0}^{N/2} x^{2n}/n!$ and the remainder term $o(x^N)$ have been separated. Substituting this expansion into the identity and regrouping terms, we obtain:
\begin{equation}
	q_N(x) g(x) = 1 - o(x^N) g(x) = 1 + o(x^N), \label{InverseTaylor}
\end{equation}
where we have used the fact that $g(x) = 1 + o(x)$. The expressions~(\ref{SolutionTaylor}) and~(\ref{InverseTaylor}) coincide when $K = N$. Therefore, $E(x) = q_N(x) g(x)$ is a solution of the original system of equations. This brings us to
\begin{equation}
	E(x) = \exp (-x^2) \sum_{n = 0}^{N/2} \frac{x^{2n}}{n!},\label{HermiteFlatTopTaylor}
\end{equation}
which agrees with the result~(\ref{HermiteFlatTop}) obtained in Appendix \ref{sec:app:AnalyticalFlatTop}. Expressing monomials $x^{2n}$ in terms of a sum of Hermite polynomials~\eqref{eq:MonomialToHermite}, we obtain superposition coefficients:
\begin{equation}
    \tilde c_{2n} = \sum_{k = n}^{N/2} \frac{(2k)!}{2^{3k} k! (k-n)! (2n)!}, \; n = 0, \dots, N/2.
\end{equation}

In Appendix~\ref{sec:app:AnalyticalFlatTop} we also present the Fourier transform $\mathcal F[E(x)]$ in Eqs.~(\ref{FourierFlatTop}) and~(\ref{FourierFlatTopKummer}), asymptotics~(\ref{AsymptoticFlatTop}), (\ref{AsymptoticFourierFlatTop}) of $E(x)$ and $\mathcal F[E(x)]$ for large $N$, and the flat-top profile construction in a polar coordinate system.

So far we have only been interested in the transverse field distribution $E(x)$ in the plane $z = 0$. The decomposition \eqref{HGsum} with the same coefficients $\tilde c_n$ is valid if we add dependence on the longitudinal coordinate $z$ to the Hermite--Gaussian mode $\mathrm{HG}_n(x,z)$:
\begin{widetext}
\begin{equation}
    \mathrm{HG}_n(x, z) = \frac{1}{\sqrt{1-iz}}\left(\frac{i-z}{i+z}\right)^{n/2} H_n \left( \frac{\sqrt{2} x}{\sqrt{1+z^2}} \right) \exp \left( -\frac{x^2}{1-iz} \right). \label{eq:HGDefinition}
\end{equation}
\end{widetext}
Here, as before, we measure transverse coordinates $x$ and $y$ in units of the Gaussian-beam waist $w_0$ and longitudinal coordinate $z$ in units of the Rayleigh length $z_0$. Taking the product over $x$ and $y$ axes, the field $\mathbf{E}_{NM}(x, y, z)$ in all space is given by
\begin{equation}
    \mathbf{E}_{NM}(x, y, z) = E_{NM}(x, y, z) \exp \left(-i \frac{2z_0^2}{w_0^2} z \right),
\end{equation}
where
\begin{equation}
    E_{NM}(x, y, z) = E_N(x, z) E_M(y, z) \label{Longitudinal}
\end{equation}
is a slowly varying part of the field. Here $N$ and $M$ are the flat-top beam orders along the $x$ and $y$ axes, respectively. Both $E_M$ and $E_N$ can be represented as superpositions. As an example, for $E_N$:
\begin{equation}
    E_N(x, z) = \sum_{n=0}^N \tilde c_n \mathrm{HG}_n(x, z). \label{eq:HGDecompositionLongitudinal}
\end{equation}

The phase factor in the expression for $\mathbf{E}_{NM}$ is the usual $\exp(-ikz)$, where the wavenumber $k$ is expressed via $w_0$ and $z_0$. In the following, we will mainly work with $E_{NM}$ instead of the total field $\mathbf{E}_{NM}$.
\begin{figure*}[!htbp]
\includegraphics[width=0.96\textwidth]{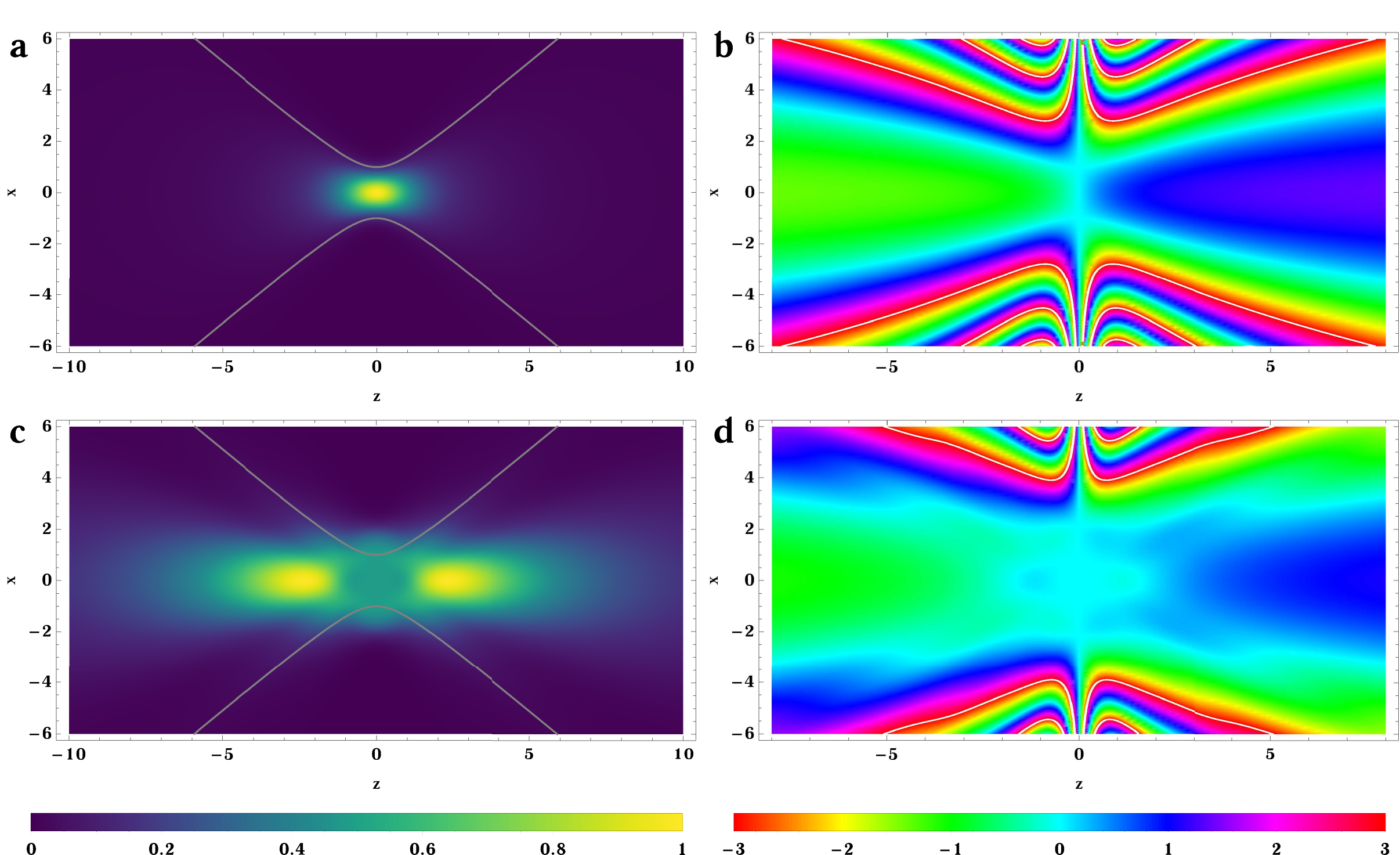}
\caption{\label{FlattopZProp} Longitudinal intensity and phase profiles for a Gaussian beam (a, b) and the flat-top beam of $x$- and $y$-axis orders $N = M = 8$ (c, d). Phase profiles do not include the factor $\exp(-ikz)$. Intensities are normalized to their respective peak values; phases are given in radians. The intensity panels also include the analytical Gaussian-beam waist for reference (gray lines).}
\end{figure*}

The longitudinal intensity and phase cross sections of the flat-top beam $E_{88}(x, 0, z)$ are shown in Fig.~\ref{FlattopZProp}. For comparison, the same distributions for a Gaussian beam $\mathrm{HG}_{00}$ are also presented.


The dependence of $E_{NM}(x, y, z)$ on $z$ is rather involved, so in Appendix~\ref{sec:app:TaylorExpansion} we present the Taylor series expansion of the longitudinal profile of such a beam and show that, e.\,g., the lowest nonvanishing power in this expansion is $N/2+1$~\eqref{eq:FlatTopTaylor} for the case $N=M$.
\section{Hologram Generation \label{sec:hologram}}
General iterative algorithms for computing phase holograms are well known. They may be broadly divided into Fourier-transform-based methods~\cite{Wu_SciRep2015} and optimization-based approaches~\cite{Bowman_OptExpress2017}. These algorithms require the specification of an initial field (in our case, a Gaussian beam) and a target beam profile in the far-field diffraction plane. The main advantages of these methods are their high flexibility, since both the input and target fields may be chosen arbitrarily, and the high theoretical accuracy of the resulting beams. Their drawbacks include long computation times and strong sensitivity to experimental imperfections. The diffraction efficiency of the computed holograms can vary significantly depending on the specified fields. \par
Knowing the field distribution in the plane $z=0$, we construct the phase mask to be applied to the SLM using the method described in \cite{Davis:1999sfk,Bolduc:2013icd}. This method is based on representing the beam reflected from the SLM as the inverse Fourier transform of the desired field distribution at $z=0$. In addition, by superimposing a blazed grating hologram, the target beam is directed into the first diffraction order, where it can be separated from the parasitic radiation caused by the unmodulated reflection from the SLM surface. \par
According to \cite{Bolduc:2013icd}, the beam reflected from the SLM can be represented as 
\[
E_{out}=\tilde A(n,m) \cdot e^{i\Phi(n,m)}=\mathcal{F}^{-1}\left[ E_{NM} \right],
\]
$E_{in}$ is assumed to be a Gaussian beam. Then the phase mask applied to the SLM is calculated as follows:
\begin{eqnarray}
    \Psi(m,n)&=&\mathcal{M}(m,n) \times \nonumber\\
    &\times& \mathrm{Mod}(\mathcal{F}(m,n)+2 \pi m/ \Lambda,\ 2\pi),
\end{eqnarray}
where 
\begin{equation}
    \mathcal{M}=1+\frac{1}{\pi}\mathrm{sinc}^{-1}(A)
\end{equation}
is a normalized bounded positive function of amplitude and
\begin{equation}
    \mathcal{F}=\Phi-\pi \mathcal{M}
\end{equation}
is an analytical function
of the amplitude and phase profiles of the desired field. \par

\begin{figure}[!htbp]
\includegraphics[height=0.22\textwidth]{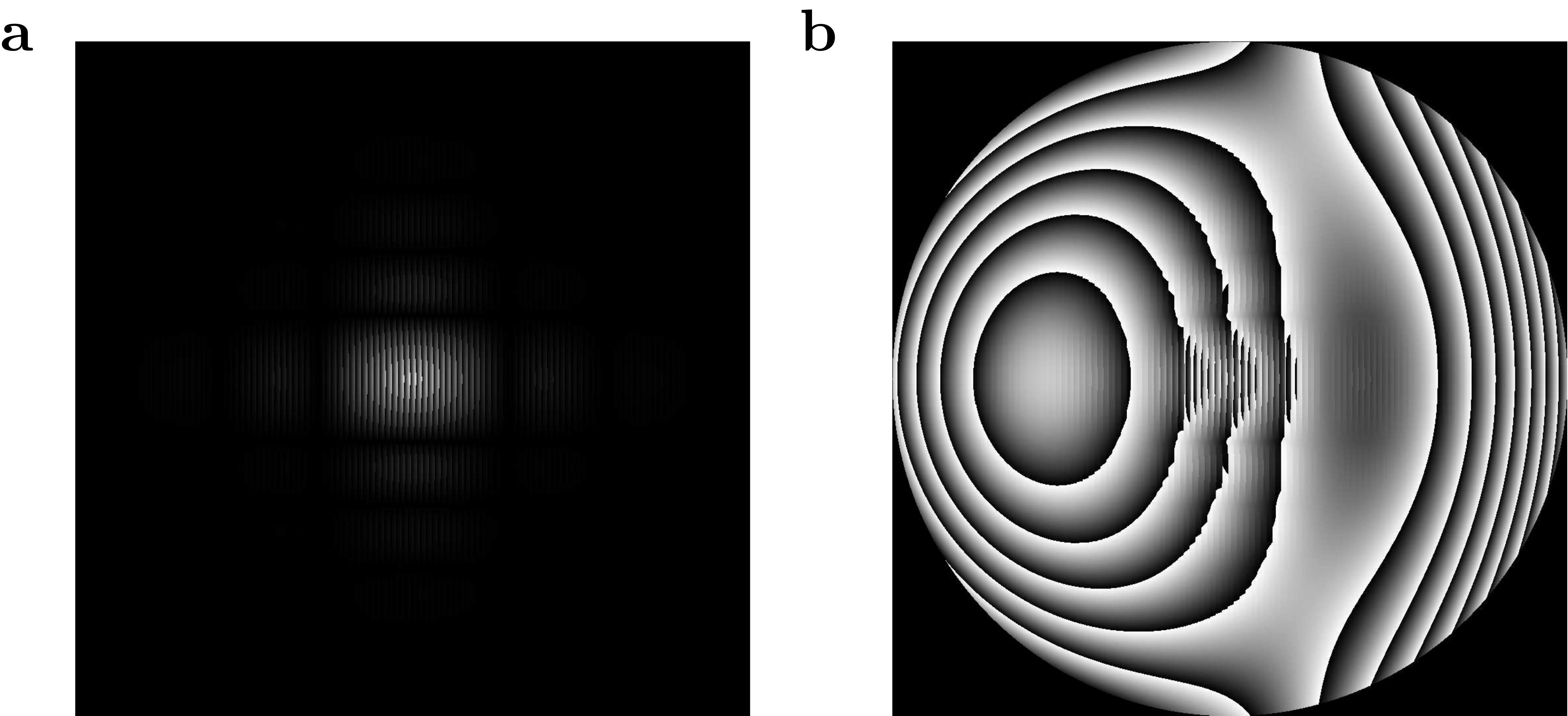}
\caption{\label{Hologram} Representative examples of holograms generated for flat-top beam preparation using an SLM. (a) Without aberration correction. (b) Superimposed with the hologram for aberration correction.}
\end{figure}

A representative example of such a hologram is shown in Fig. \ref{Hologram}a. \par
Then, we compensate for optical aberrations by adding corrective holograms constructed from the corresponding Zernike polynomials. In particular, we correct for vertical astigmatism ($Z_2^2$) and horizontal coma ($Z_3^1$) according to the equations given in \cite{born1980} by using a linear combination of holograms ($a_2 Z_2^2+a_3 Z_3^1$) as described in \cite{Sorimoto:2010faa}. To find the coefficients $a_2$ and $a_3$, we diverted a small fraction of the radiation using an amplitude beam splitter to a laser beam profiler positioned after a focusing lens. This allowed us to observe the far-field intensity distribution. The coefficients were then optimized to maximize the uniformity of the flat-top profile while minimizing the parasitic radiation around it. Figure~\ref{Hologram}b shows an example of the resulting hologram.
\section{Numerical model \label{sec:model}}
We numerically investigate the Rydberg-gate performance achievable with the flat-top beam considered here. Our model incorporates the dominant physical error sources, including intermediate-state decay, Rydberg-state decay, and decoherence induced by atomic thermal motion. The model is implemented as the Julia package NeutralAtoms.jl \cite{NeutralAtoms} and includes an additional module that accounts for laser phase noise following \cite{Jiang_2023}.

\begin{figure}[!htbp]
\includegraphics[height=0.22\textwidth]{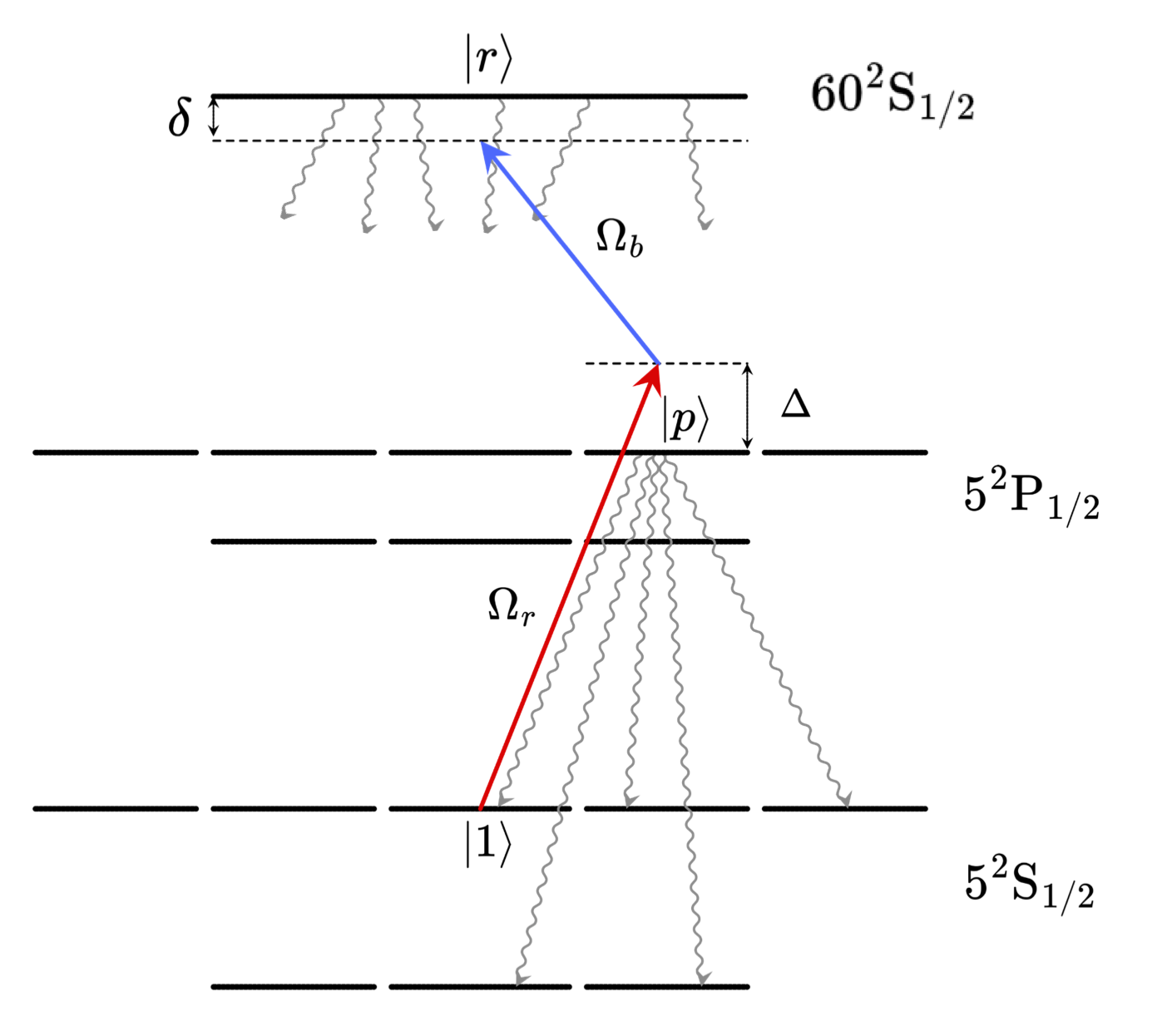}
\includegraphics[height=0.22\textwidth]{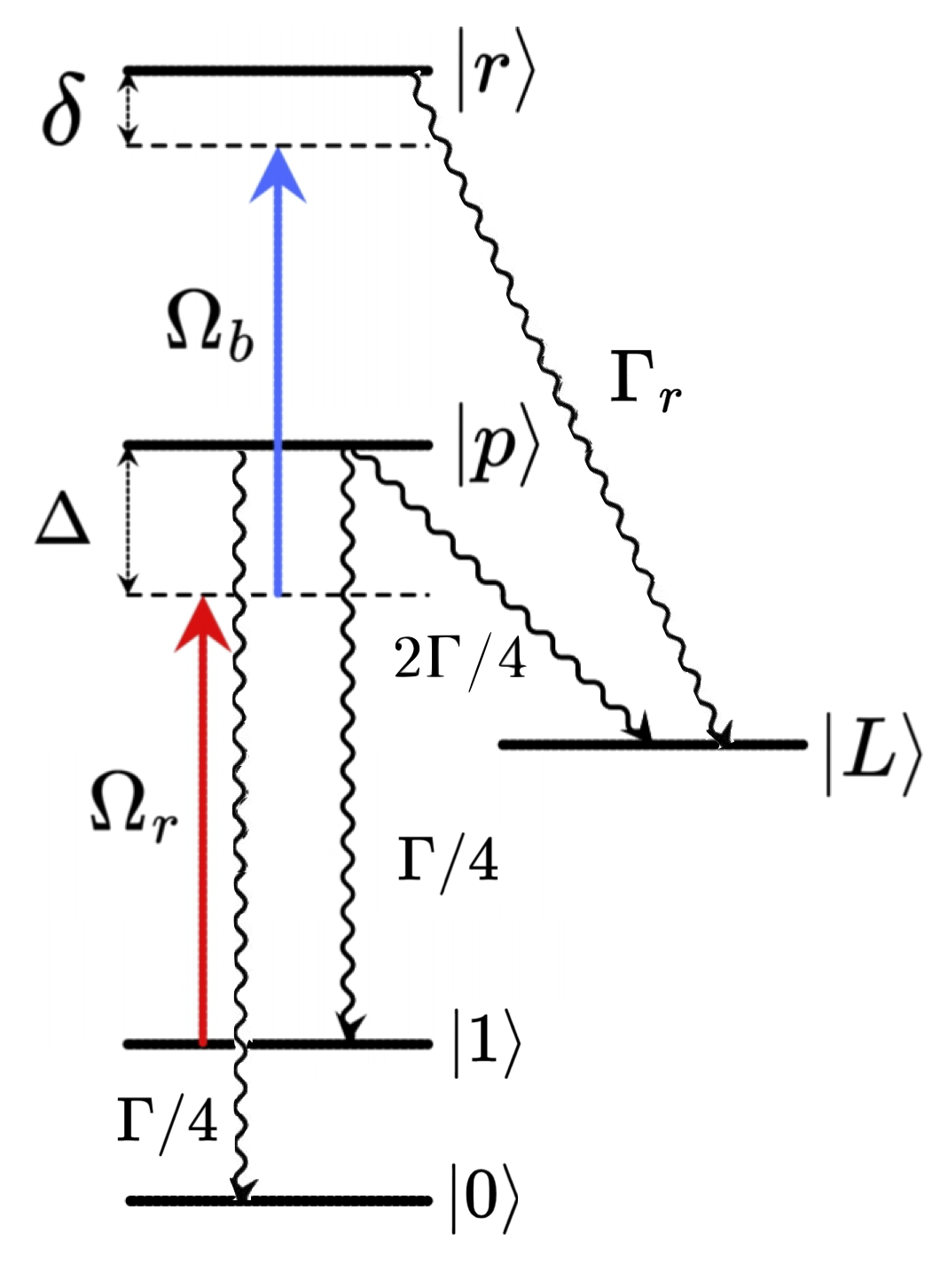}
\caption{\label{AtomicLevels} Effective 5-level system. (a) Full hyperfine structure of $5^{2}\textrm{S}_{1/2}$ and $5^{2}\textrm{P}_{1/2}$. (b) Reduced atomic levels with effective decay level $\ket{L}$.}
\end{figure}

Both the intermediate state $\ket{p}$ and the Rydberg state $\ket{r}$ decay via multiple channels. To keep the model tractable while preserving the dominant loss mechanisms, we introduce an effective level $\ket{L}$, similar to that in \cite{de_L_s_leuc_2018}, which accumulates population decaying into non-target states. The reduction from the full atomic model to the effective model is shown in Fig.~\ref{AtomicLevels}, with branching ratios computed using the ARC library \cite{SIBALIC2017319}. \par
Given this effective level structure, we define the Hamiltonian and the jump operators relevant to CZ-gate implementation:
\begin{subequations}
\begin{eqnarray}
    H_{i} = & \frac{\Omega_{r,i}}{2}\ket{1_i}\bra{p_i} + \frac{\Omega^*_{r,i}}{2}\ket{p_i}\bra{1_i} \nonumber\\
    + &\frac{\Omega_{b,i}}{2}\ket{p_i}\bra{r_i} + \frac{\Omega_{b,i}^*}{2}\ket{r_i}\bra{p_i} \nonumber\\
    -&\Delta_{i} \ket{p_i}\bra{p_i} -\delta_{i} \ket{r_i}\bra{r_i}.
\end{eqnarray}
\begin{equation}
    H = H_{1} \otimes I + I \otimes H_2 + V\ket{r_1}\bra{r_1}\otimes \ket{r_2}\bra{r_2}.
\end{equation}
\begin{eqnarray}
    J_{0p}=\sqrt{\Gamma/4}\ket{1}\bra{p} ,\; J_{1p}=\sqrt{\Gamma/4}\ket{1}\bra{p}, \nonumber\\
    J_{Lp}=\sqrt{\Gamma/2}\ket{L}\bra{p}, \; J_{Lr}=\sqrt{\Gamma_r}\ket{L}\bra{r}.
\end{eqnarray}
\end{subequations}

The effective system dynamics are simulated by solving the time-dependent master equation in Lindblad form using QuantumOptics.jl \cite{kramer2018quantumoptics}. As discussed below, the master equation becomes time-dependent due to atomic dynamics:
\begin{equation}
    \dot{\rho} = -i[H, \rho] + \sum_{i}\left(J_{i}\rho J_{i}^{\dagger} - \frac{1}{2}\left\{J_i^\dagger J_{i}, \rho\right\}\right).
\end{equation}

Thermal motion is incorporated in several steps. First, atomic positions and velocities are sampled from a Boltzmann distribution in a harmonic trap potential at temperature $T$. The optical-tweezer potential with beam waist radius $w_0$ and Rayleigh length $z_0$ is
\begin{eqnarray}
    U = & U_0\left(1 - \frac{w_0^2}{w(z)^2}\exp\left(-\frac{2(x^2+y^2)}{w(z)^2}\right)\right), \nonumber\\
    & w(z) = w_0\sqrt{1+(z/z_0)^2}.
\end{eqnarray}

Near the trap center, this potential is approximated by a harmonic trap with radial and axial frequencies $\omega_r$ and $\omega_z$, respectively:
\begin{equation}
    \omega_r = \frac{1}{w_0}\sqrt{\frac{4U_0}{m}}, \; \omega_z = \frac{1}{z_0}\sqrt{\frac{2U_0}{m}}.
\end{equation}

Additionally, we implemented sampling from the exact optical-tweezer potential using the Metropolis--Hastings algorithm and observed no significant deviation from the harmonic approximation for temperatures up to 100$\upmu K$. Because the Boltzmann distribution in a Gaussian potential is not normalizable over infinite space, a finite spatial cutoff is introduced during sampling.

After sampling the initial phase-space coordinates, atomic trajectories are propagated assuming the trapping potential is switched off during Rydberg excitation to avoid anti-trapping effects. These trajectories yield time-dependent Hamiltonian parameters $\Delta_i(t), \; \delta_i(t), \; \Omega_{r,i}(t), \; \Omega_{b,i}(t), \; V(t)$. The master equation is then solved for each trajectory and the resulting density matrices are averaged in a Monte Carlo fashion. Alternatively, one can use the Monte Carlo wave function (MCWF) \cite{Molmer:93} method to avoid solving the computationally expensive master equation. \par
In the experiment, we measure Rydberg-excitation Rabi oscillations on the target atoms and record observables characterizing crosstalk on neighboring atoms. We then use numerical modeling to infer the set of experimental parameters governing crosstalk, including those that are not directly accessible because of the constraints imposed by the vacuum-chamber geometry. These results are presented in Sec.~\ref{sec:results}. \par
In Appendix~\ref{sec:app:params} we present additional information about measurements of the parameters useid in numerical model (atom temperature, parameters of the optical traps and reconstruction of the laser beams acting on atoms).
\section{Experimental setup \label{sec:setup}}
\begin{figure*}[b]
\includegraphics[width=0.97\textwidth]{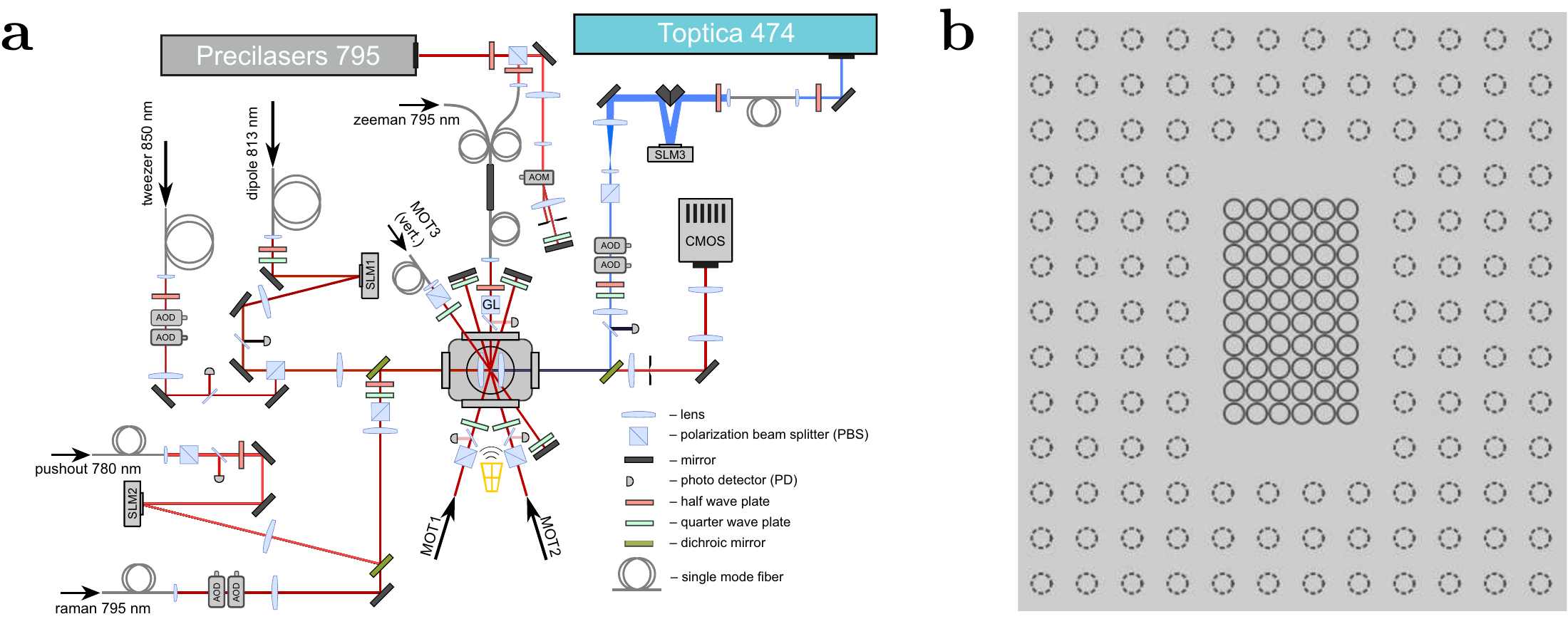}
\caption{\label{RydbergExScheme} (a) Optical scheme of the experimental setup. (b) Computational zone (shown centered; when forming the computational array, either the five leftmost or the five rightmost columns are used) and storage zone.}
\end{figure*}

The experimental setup is shown in Fig.~\ref{RydbergExScheme}a. All manipulations with the $^{87}$Rb atoms are carried out inside a vacuum chamber, where a 45 L/s ion pump maintains a pressure of $1.2\times10^{-10}$~mbar. Two dispensers supply rubidium vapor to the operating region. \par
$^{87}$Rb atoms are first captured in a magneto-optical trap (MOT) formed by three pairs of laser beams containing cooling light at 780~nm and repumping light at 795~nm. The cooling beams are red-detuned by $4\Gamma$ from the $5S_{1/2} F=2\rightarrow5P_{3/2} F=3$ transition and have a beam radius of 0.96 mm (here and throughout, beam waists $w_0$ and radii $w(z)$ are specified as the $1/e^2$ intensity radii). Two electromagnetic coils in an anti-Helmholtz configuration generate a field gradient of 12.1~G/cm at the point where the magnetic field is zero. \par
Single atoms are subsequently loaded from the magneto-optical trap into optical dipole traps in a storage zone surrounding the computational array. The storage zone contains 128 atoms trapped at a spacing of 7.2~$\upmu$m (Fig.~\ref{RydbergExScheme}b). The dipole traps are formed by an 813 nm laser beam; each trap has a beam waist of 1.4~$\upmu$m and a power of 2.5~mW. The computational array consists of $5\times10$ sites with a spacing of 3.6~$\upmu$m between neighboring atoms. It is filled with atoms from the storage zone using optical tweezers (a tweezer beam has a wavelength of 852 nm) controlled by acousto-optical beam deflectors and arranged according to a Hungarian algorithm. \par
A pair of LightPath 355561 lenses with an effective focal length of 10~mm and a working distance of 7~mm is installed inside the vacuum chamber. The left lens provides tight focusing of the dipole-trap beam. The right lens focuses the second-stage Rydberg-excitation beam, which propagates in the opposite direction to the dipole-trap and tweezer beams, onto the atoms and also directs the imaging signal toward the back-illuminated sCMOS camera.
\begin{figure*}[!htbp]
\includegraphics[width=0.95\textwidth]{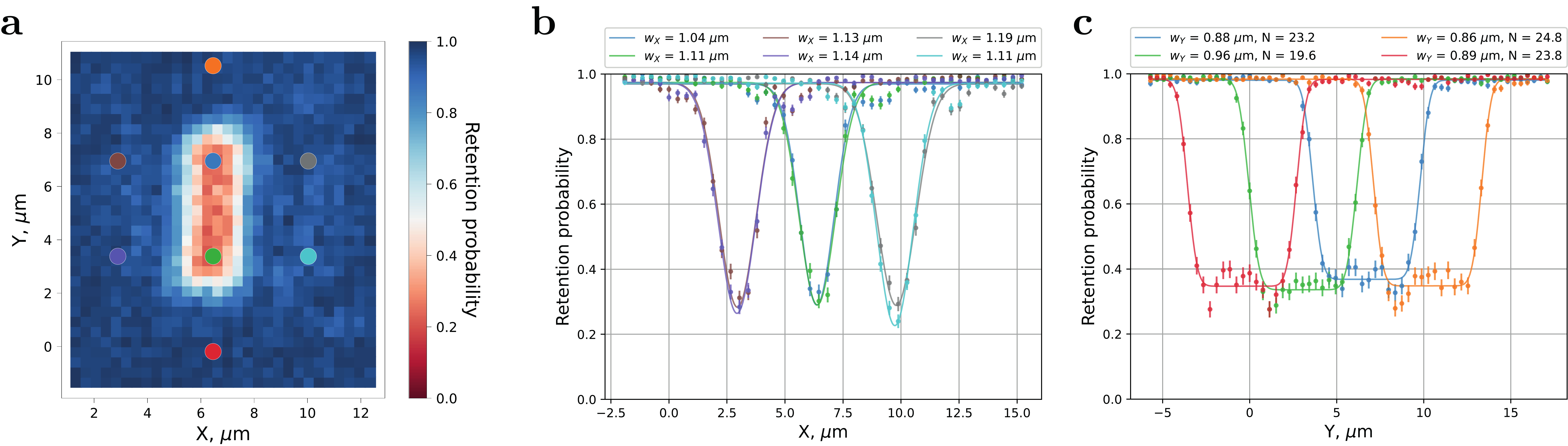}
\caption{\label{Profiles} (a) Atom ionization density map. The addressed atoms are shown in blue and green. Other colors correspond to the non-addressed atoms. (b) Transverse profile of the beam along the x-axis approximated by Gaussian. (c) Transverse profile of the beam along the y-axis approximated by a flat-top profile.}
\end{figure*}
\begin{figure*}[t]
\includegraphics[width=0.98\textwidth]{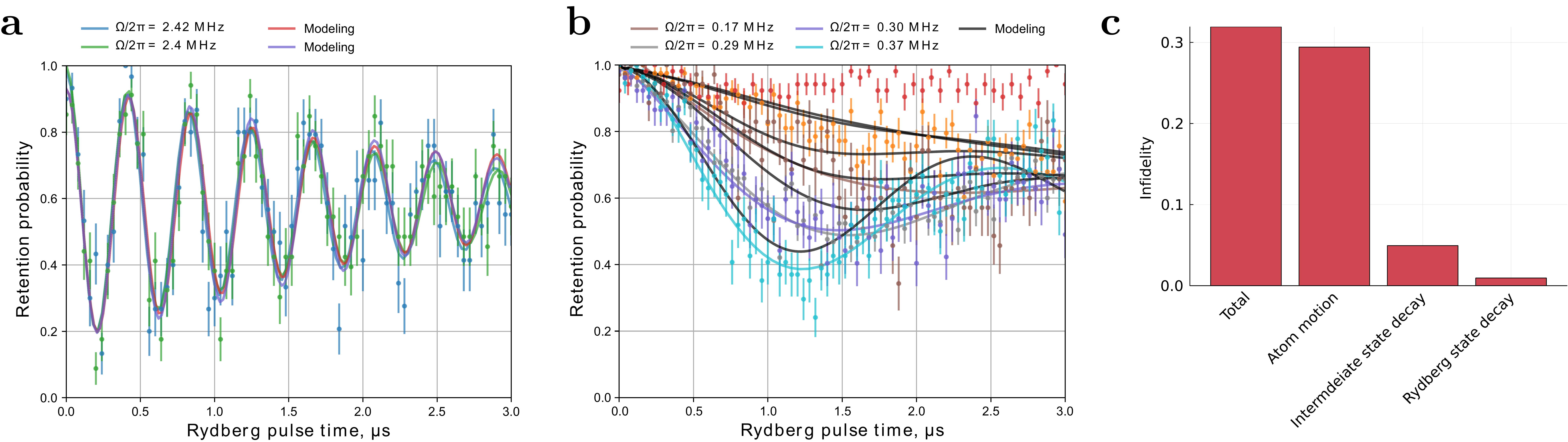}
\caption{\label{Results} Modeling and experimental results. (a) Rydberg Rabi oscillations on addressed atoms. (b) Rydberg Rabi oscillations on nonaddressed atoms. (c) CZ gate error budget.}
\end{figure*}

We apply polarization-gradient cooling to the atoms in the optical dipole traps. It is implemented using the magneto-optical trap beams after the MOT magnetic field has been switched off; at this stage, the 780 nm light is reduced in power and detuned farther from resonance, while the 795 nm repumping light is kept at the same frequency as during MOT operation. As a result, the temperature of the atoms in individual traps is reduced to slightly below 100~$\upmu$K. \par
We implement the CZ gate proposed in Ref.~\cite{Levine:2019zfq} on selected pairs of atoms in the target array using two-photon Rydberg excitation to $\ket{r}=\ket{60S_{1/2}, m_j=1/2}$. The qubit states are encoded in the hyperfine ground-state sublevels $\ket{0}=\ket{5S_{1/2}, F=1, m_F=0}$ and $\ket{1}=\ket{5S_{1/2}, F=2, m_F=0}$. We prepare the first of these states by Zeeman pumping using the method described in Ref.~\cite{Levine:2019zfq}. Then we rotate the atomic states to $\ket{1}$ using microwave radiation resonant with the 6.834682~GHz hyperfine transition. \par
The first-stage Rydberg excitation beam has a wavelength $\lambda_1 = 795$~nm and drives the transition from the state $\ket{1}$ to the intermediate state $\ket{p} = \ket{5P_{1/2}, F =~1, m_F = 0}$ with a blue detuning $\Delta/2 \pi=870$~MHz from resonance (Fig.~\ref{Flattop}a). This beam is global, with a power of 100~mW and a beam waist of 0.57~mm (corresponding to a peak intensity of approximately 196~mW/mm$^2$). The atoms are then excited to $\ket{r}$ by a tightly focused second-stage beam at $\lambda_2 = 474$~nm, which provides site-selective addressing (peak intensity $8.6\cdot10^6$~mW/mm$^2$). \par
The second-stage beam must provide a uniform intensity distribution over the region where the addressed atom pair is located while avoiding interaction with neighboring atoms of the computational register to prevent decoherence of their states. In our setup, this requirement is met by shaping the beam into a flat-top profile using an SLM with a pixel size of 12.5 $\upmu$m and a resolution of $1272\times 1024$. The beam incident on the SLM has a diameter of 6~mm and passes through a pair of steering mirrors that set the beam at a small angle of incidence on the SLM. This angle cannot be made arbitrarily small, since the reflected beam must pass aside the second steering mirror without clipping. \par
Following the method described in Sections\,\ref{sec:theory}--\ref{sec:hologram}, we obtain the phase mask to be applied to the SLM that converts a Gaussian beam into a flat-top profile. After reflection from the SLM, the beam passes through a demagnifying telescope, two orthogonally oriented acousto-optic deflectors (AODs) used for addressing, and a lens that compensates chromatic aberrations. The light is then coupled into the vacuum chamber via a dichroic mirror and reaches the plane of the dipole traps, where it acquires a nearly rectangular intensity profile.
\section{Results \label{sec:results}}
To determine the characteristics of the resulting profile, we measure the atomic ionization probability while simultaneously illuminating the atoms with the second-stage excitation beam and a 780~nm beam, which populates the 5P$_{3/2}$ state. Figure~\ref{Profiles}a shows the density map obtained by scanning the beam along the $x$ and $y$ axes using the AODs. The color scale represents the probability that an atom remains trapped. The frequencies of the sinusoidal signals applied to the AODs for beam deflection have been converted into the corresponding geometric coordinates in the plane of the dipole-trap array. Figure~\ref{Profiles}b shows the cross section along the $x$ axis, whereas Fig.~\ref{Profiles}c shows the cross section along the $y$ axis. Along the $x$ axis, the beam has a Gaussian intensity distribution with a waist of $w_x=1.1\ \upmu$m; along the $y$ axis, the beam exhibits a flat-top profile with a half-width of 3.0~$\upmu$m. \par
To assess the quality of the two-qubit operations implemented in our system, Rabi oscillations of the Rydberg transition were measured on two addressed qubits. The results are shown in Fig.~\ref{Results}a-b. Well-resolved oscillations are observed with frequencies $\Omega/2 \pi$ = 2.42~MHz and 2.40~MHz, indicating an intensity uniformity exceeding 99\%. The corresponding average Rabi frequency is $\bar \Omega/2 \pi=2.41(0.01)$~MHz, from which a CZ gate time of 287~ns is obtained. \par
We simulated these oscillations in the NeutralAtoms.jl package in two different ways. Figure~\ref{Results}a presents the simulation for the case in which $\bar \Omega/2\pi = 2.41(0.01)$~MHz is taken as an experimentally measured input parameter. When the measured powers and sizes of the Rydberg-excitation beams given in Sec.~\ref{sec:setup} are used instead, the calculation gives $\bar \Omega/2\pi = 2.84$~MHz. This suggests the presence of additional experimental error sources, such as additional losses due to the vacuum-chamber windows antireflection coating, which is imperfect for 474~nm light. \par
Rabi oscillations were also measured on six neighboring atoms, which are likewise shown in Fig.~\ref{Results}b.
To characterize crosstalk, we estimate the crosstalk field experienced by the atoms by fitting the experimental data to damped oscillations:
\begin{gather}
P(t) = 1 - \dfrac{1}{2}\dfrac{\Omega_0^2}{\tilde \Omega^2} (1-e^{-\gamma t}\cos{\tilde \Omega t}), \\ \tilde \Omega = \sqrt{\Omega_0^2+\delta^2},
\end{gather}
where $\gamma$ is a decay factor and $\delta$ is the detuning of the driving field seen by the neighboring atom relative to the target atom due to spatially varying light shifts or magnetic-field gradients. In the numerical model we take into account different Stark shifts for the non-target atoms. \par
Assuming that atoms experience the same intensity of the 795 nm beam with a waist of 0.57 mm, ratio of fields from the 474 nm laser $E_1$ and $E_2$ experienced by target and non-target atoms will be equal to the ratio of two-photon Rabi frequencies:
\begin{equation}
\eta = \frac{E_2}{E_1} = \frac{\Omega_0}{\bar \Omega}. 
\label{crosstalk}
\end{equation}

For atoms located above and below the addressed ones (marked by red and orange points in Fig.~\ref{Profiles}a), the parameter $\eta$ does not exceed 2\%;  excitation of these atoms is negligible. For atoms located to the left and right of the entangled pair, the excitation cannot be considered small: they undergo off-resonant excitation with frequencies $ \Omega_0/2 \pi$ ranging from 0.11 to 0.29~MHz. The maximum value of the crosstalk parameter is 12\%. These values can be reduced by an appropriate correction of the phase mask.  \par
Fig.~\ref{Results}c shows error budget for CZ gate implementation in our experimental setup. This shows that the dominant contribution to the imperfection comes from the residual thermal motion of the atoms in the dipole traps.
\section{Conclusion \label{sec:conclusion}}
The high-quality shaping of the addressing beam to produce a flat-top intensity distribution in the region occupied by the target atoms provides nearly uniform coupling to the selected qubits while suppressing unwanted interaction with neighboring atoms of the computational register. \par
In this work, we developed a method for generating such beams based on representing the desired flat-top profile as a superposition of low-order even Hermite–Gaussian modes with vanishing derivatives at the beam center. We derived explicit analytical expressions for the expansion coefficients, analyzed the asymptotic behavior of the resulting profiles and their Fourier transforms, and proposed an extension to the radially symmetric case using Laguerre–Gaussian modes. \par
Analysis of the beam propagation near the waist shows that the beam retains a certain phase flatness along the propagation direction $z$. The phase varies more slowly than for a Gaussian beam. \par
We implemented the method experimentally using a spatial light modulator that encodes a phase hologram that transforms an incident Gaussian beam into a flat-top profile in the focal plane. Compensation of optical aberrations using Zernike polynomials allowed us to obtain the required intensity distribution in the plane of the atomic traps. We characterized the beam by scanning it with acousto-optic deflectors and measuring the atomic ionization probability under controlled illumination, which enabled us to determine its geometric parameters. \par
In addition, we simulated Rydberg-excitation Rabi oscillations using our Julia-based library. From this modeling, we estimated the crosstalk affecting neighboring atoms and found that the damping of the oscillations on target atoms is caused predominantly by significant atomic thermal motion. \par
The proposed approach provides a practical route to spatially selective excitation in large arrays of neutral atoms and is directly applicable to scalable neutral-atom quantum computing platforms, where uniform control of selected qubits with minimal crosstalk to neighboring ones is essential.

\begin{acknowledgments}
The results presented in Sections~II--VI and Appendices~A,~C were supported by Rosatom in the framework of the Roadmap for Quantum computing (Contract No. 868-1.3-15/15-2021 dated October 5, 2021 and Contract No. P2154 dated November 24, 2021).

The results presented in Appendix B were supported by Rosatom in the framework of the Roadmap for Quantum computing (Contract No. 868/1653-D dated August 21, 2025). 
\end{acknowledgments}
\appendix

\section{Analytical aspects of flat-top beams\label{sec:app:AnalyticalFlatTop}}
In the main text, we presented an overview and the principal results of our model for flat-top beam formation. Here we provide detailed analytical derivations of the expansion coefficients in the Hermite–Gaussian basis, together with several additional analyses related to the generation of such beams.

\subsection{Direct analytical solution for the Hermite–Gaussian expansion coefficients\label{sec:HGFlatTopSolution}}

It is natural to require the profile $E(x)$ to be symmetric, i.e., $E(x) = E(-x)$. This implies that $c_n = 0$ for odd $n$ and $d^k E(0)/dx^k = 0$ for odd $k$. Accordingly, it suffices to consider only even values of $N$ and $K$ in what follows. From~(\ref{Normalization}) we obtain $c_0 = 1$. Substituting~(\ref{SuperposPoly}) into (\ref{ZeroDeriv}), we obtain:
\begin{equation}
	\left. \sum_{n = 0}^{N/2} c_{2n} \frac{d^{2k}(x^{2n} \exp(-x^2))}{dx^{2k}}\right|_{x=0} = 0, \ k = 1, \dots, K/2.
\end{equation}
For $n > k$, the derivative at $x = 0$ is clearly zero. For $n \le k$, we expand the derivative using the Leibniz rule:
\begin{widetext}
\begin{eqnarray}
	\frac{d^{2k}(x^{2n} \exp(-x^2))}{dx^{2k}} &&= \sum_{j=0}^{2k} \frac{(2k)!}{(2k-j)! j!} \frac{d^jx^{2n}}{dx^j} \frac{d^{2k-j}\exp(-x^2)}{dx^{2k-j}} = \notag\\ &&= \sum_{j=0}^{2k} \frac{(2k)!}{(2k-j)! j!} \frac{d^j x^{2n}}{dx^j} (-1)^{2k-j} \exp(-x^2) H_{2k-j}(x).
\end{eqnarray}
\end{widetext}
Using $\left.d^j(x^{2n})/dx^j\right|_{x=0} = (2n)! \delta^j_{2n}$ and $H_{2m}(0) = (-1)^{m} (2m)! / m!$, we obtain:
\begin{gather}
	\left.\frac{d^{2k}(x^{2n} \exp(-x^2))}{dx^{2k}} \right|_{x=0} \notag = \\ \begin{cases}
		(-1)^{k-n} \frac{(2k)!}{(k-n)!}, & n \le k, k = 1, \dots, K/2, \\
		0, & n > k.
	\end{cases} \label{SystemMatrix}
\end{gather}

From (\ref{SystemMatrix}), it is evident that the system matrix of~(\ref{ZeroDeriv}--\ref{Normalization}) is triangular. Consequently, the system is consistent only when $K \le N$, with the maximum value $K = N$. In this case, (\ref{ZeroDeriv}--\ref{Normalization}) reduces to the equivalent system:
\begin{gather}
	c_0 = 1, \\
	\sum_{n = 0}^{k} c_{2n} (-1)^{k-n} \frac{(2k)!}{(k-n)!} = 0, \quad k = 1, \dots, N/2.
\end{gather}
The solution is given by $c_{2n} = 1/n!$ and $c_{2n+1} = 0$. This follows from the identity below, obtained using the binomial theorem:
\begin{equation}
	0 \equiv \frac{(2k)!}{k!} (1-1)^k = \sum_{n=0}^k \frac{1}{n!} (-1)^{k-n}\frac{(2k)!}{(k-n)!}.
\end{equation}

Upon substituting this solution, the flat-top beam profile~(\ref{SuperposPoly}) takes the form:
\begin{equation}
	E(x) = \exp (-x^2) \sum_{n = 0}^{N/2} \frac{x^{2n}}{n!}. \label{HermiteFlatTop}
\end{equation}

This profile can be written as an explicit expansion in Hermite–Gaussian modes~(\ref{SuperposHermite}) by determining the coefficients~$\tilde c_n$. To this end, we express $x^{2n}$ in terms of $H_{2k}(\sqrt{2} x)$~\cite{DLMF}:
\begin{equation}
	x^{2n} = \frac{(2n)!}{2^{3n}} \sum_{j=0}^{n} \frac{H_{2n-2j}(\sqrt{2} x)}{j! (2n-2j)!}. \label{eq:MonomialToHermite}
\end{equation}
Substituting the expansion into~(\ref{HermiteFlatTop}) and collecting terms with the same Hermite polynomial $H_{2k}$ (where $2k := 2n-2j$), followed by interchanging the order of summation over $n$ and $k$, we obtain:
\begin{eqnarray}
	E(x) &=& \exp (-x^2) \sum_{k = 0}^{N/2} \left( \sum_{n = k}^{N/2} \frac{(2n)!}{2^{3n} n! (n-k)! (2k)!} \right) \times \nonumber\\  & &\times H_{2k}(\sqrt{2} x). \label{HermiteFlatTopCoeffs}
\end{eqnarray}
The expression in parentheses is precisely the desired coefficient $\tilde c_{2k}$. \par
The sum~(\ref{HermiteFlatTop}) can be written exactly in compact form using a known property of the regularized incomplete gamma function $Q(a, z)$ \cite{SpecFuncBook_1979}:
\begin{equation}
	E(x) = Q(N/2 + 1, x^2), \label{GammaFlatTop}
\end{equation}
where
\begin{eqnarray}
	Q(a, z) &:=& \frac{\Gamma(a, z)}{\Gamma(a)}, \quad \Gamma(a) := \Gamma(a, 0),\nonumber\\ \Gamma(a, z) &:=& \int_z^\infty t^{a-1} e^{-t} \,dt,
\end{eqnarray}
where $\Gamma(a, z)$ is the incomplete gamma function. \par
Formally, the expression~(\ref{GammaFlatTop}) is also defined for real, noninteger $N$. In this case, it may be viewed as an interpolation to arbitrary $N$, which is convenient for analytical studies.

\subsection{Asymptotics}
Formula~(\ref{GammaFlatTop}) is more convenient than~(\ref{HermiteFlatTop}) for analyzing the asymptotic behavior as $N \to \infty$. Using Theorems 1.1 and 1.2 of Ref.~\cite{Nemes_MathComp2019}, we find that, to leading order, $E(x)$ converges to the complementary error function $\erfc$:
\begin{equation}
	E(x) = \frac{1}{2} \erfc\left(\sqrt{2}|x|-\sqrt{N + 4/3} \right) + O(1/\sqrt{N}), \label{AsymptoticFlatTop}
\end{equation}
In practice, the approximation is accurate already for $N \gtrsim 10$. For example, at $N = 10$ the error does not exceed $0.014$ for all $x$. \par
The asymptotic behaviour implies that the full width at half maximum of the flat beam is approximately $\sqrt{2N + 8/3}$, while the width of the transition region, where the intensity decreases from 1 to 0, approaches a constant value~\footnote{Note that it is proportional to the Hermite--Gaussian mode waist $w_0$, since $x$ is measured in units of $w_0$.}.

\subsection{Fourier transform}
For certain hologram-computation algorithms (e.g., the method of Ref.~\cite{Bolduc:2013icd}), the required input is not the desired field profile $E(x)$ itself but its Fourier transform\footnote{In what follows, we omit the argument of the Fourier transform, assuming it is denoted by the same symbol $x$ instead of $t$.} $\mathcal F[E(x)](t) = \frac{1}{\sqrt{2\pi}} \int_{-\infty}^{\infty} E(x) \exp(i x t)\,dx$. Since the Fourier transform is linear, computing $\mathcal F[E(x)]$ from~(\ref{HermiteFlatTop}) reduces to evaluating $\mathcal F[x^{2n} e^{-x^2}]$. For any sufficiently fast-decaying function $g(x)$, the following identity holds:
\begin{eqnarray}
	&&\frac{d\mathcal F[g(x)]}{dx} = \mathcal F[i x g(x)] \Rightarrow \nonumber\\ \Rightarrow &&\frac{d^{2n} \mathcal F[g(x)]}{dx^{2n}} = (-1)^n \mathcal F[x^{2n} g(x)].
\end{eqnarray}
Setting $g(x) = \exp(-x^2)$ and using $\mathcal F[\exp(-x^2)] = \exp(-x^2/4)/\sqrt{2}$, we obtain
\begin{eqnarray}
	\mathcal F[x^{2n} \exp(-x^2)] = \frac{(-1)^n}{\sqrt{2}} \frac{d^{2n}\exp(-\frac{x^2}{4})}{dx^{2n}} = \nonumber\\ = \frac{(-1)^n}{2^{2n}\sqrt{2}} \exp \left( -\frac{x^2}{4} \right) H_{2n} \left( \frac{x}{2} \right).
\end{eqnarray}
It follows that the Fourier transform of~(\ref{HermiteFlatTop}) is given by
\begin{eqnarray}
	\mathcal F[E(x)] &=& \frac{S_{N/2}}{\sqrt{2}} \exp \left( -\frac{x^2}{4} \right), \nonumber\\ S_{N/2} &=& \sum_{n=0}^{N/2} \frac{(-1)^n}{2^{2n} n!} H_{2n} \left( \frac{x}{2} \right). \label{FourierFlatTopSum}
\end{eqnarray}

The sum $S_{N/2}$ can be further simplified using known recurrence relations for Hermite polynomials~\cite{SpecFuncBook_1979}. Using these relations, we can express $H_{2n}(x/2)$:
\begin{equation}
	x H_{2n}(x/2) = 4 n H_{2n-1}(x/2) + H_{2n+1}(x/2). \label{RecursiveHermite}
\end{equation}
Substituting~(\ref{RecursiveHermite}) into~(\ref{FourierFlatTopSum}), we obtain:
\begin{eqnarray}
	x S_{N/2} &=& \sum_{n=1}^{N/2} \frac{(-1)^n 4n H_{2n-1}(x/2)}{2^{2n} n!} + \nonumber\\ &+& \sum_{n=0}^{N/2} \frac{(-1)^n H_{2n+1}(x/2)}{2^{2n} n!}.
\end{eqnarray}
Performing the substitution $m = n - 1$ in the first sum, we obtain:
\begin{eqnarray}
	x S_{N/2} &=& -\sum_{m=0}^{N/2-1} \frac{(-1)^m H_{2m+1}(x/2)}{2^{2m} m!} + \nonumber\\ &+& \sum_{n=0}^{N/2} \frac{(-1)^n H_{2n+1}(x/2)}{2^{2n} n!}.
\end{eqnarray}
In the resulting expression, the sums nearly cancel each other, leaving only the final term of the second sum. As a result,~(\ref{FourierFlatTopSum}) takes the form:
\begin{eqnarray}
	\mathcal F[E(x)] &=& \frac{(-1)^{N/2}}{2^{N} (N/2)! \sqrt{2}} \times \nonumber \\ &\times& \frac{H_{N+1}(x/2)}{x} \exp \left( -\frac{x^2}{4} \right).
\end{eqnarray}

Formally, the expression has a removable singularity at $x = 0$, and its value should therefore be understood as the limit $x \to 0$. The ratio $H_{N+1}(x/2) / x$ is in fact a polynomial of order $N$. Using the relation between odd-order Hermite polynomials $H_{N+1}$ and the associated Laguerre polynomials $L_{N/2}^{(1/2)}$~\cite{SpecFuncBook_1979}, the expression can be rewritten without division by $x$:
\begin{equation}
	\mathcal F[E(x)] = \frac{1}{\sqrt{2}} L_{N/2}^{(1/2)}  \left( \frac{x^2}{4} \right) \exp \left( -\frac{x^2}{4} \right).
\end{equation}
In practice, it is convenient to impose the normalization $\mathcal F[E(x)](0) = 1$. In this normalization,
\begin{eqnarray}
	\mathcal F[E(x)] = \frac{(-1)^{N/2} (N/2)!}{(N+1)!} \times \nonumber\\ \times \frac{H_{N+1}(x/2)}{x} \exp\left(-\frac{x^2}{4} \right). \label{FourierFlatTop}
\end{eqnarray}

To analyze the asymptotic behavior of~(\ref{FourierFlatTop}) in the limit $N \to \infty$, we express the odd Hermite polynomial $H_{N+1}(x/2)$ in terms of Kummer's confluent hypergeometric function  $M(a, b, z)$ \cite{SpecFuncBook_1979}:
\begin{eqnarray}
	\mathcal F[E(x)] &=& M \left( -\frac{N}{2}, \frac{3}{2}, \frac{x^2}{4} \right) \exp\left(-\frac{x^2}{4} \right) \nonumber\\ &=& M \left( \frac{N+3}{2}, \frac{3}{2}, -\frac{x^2}{4} \right), \label{FourierFlatTopKummer}
\end{eqnarray}
where, in the final equality, the so-called Kummer transformations were applied. Using the asymptotic expansion of the Kummer function for large $N$~\cite{SpecFuncBook_1979}, we obtain:
\begin{eqnarray}
	\mathcal F[E(x)] &=& \sinc\left(x \sqrt{\frac{N}{2} + \frac{3}{4}}\right) \times \nonumber\\ &\times& \exp\left(-\frac{x^2}{8} \right) \left(1 + O(1/\sqrt{N})\right). \label{AsymptoticFourierFlatTop}
\end{eqnarray}
We note that, up to terms of order $O(1/\sqrt{N})$, the asymptotic form~(\ref{AsymptoticFourierFlatTop}) can be obtained directly by taking the Fourier transform of~(\ref{AsymptoticFlatTop}) and using the relation:
\begin{eqnarray}
	&& \mathcal F\left[ \exp\left(-\frac{x^2}{a^2} \right) \sinc(b x) \right] \propto \nonumber\\ \propto&& \erfc \left( \frac{a(x-b)}{2} \right) - \erfc \left( \frac{a(x+b)}{2} \right).
\end{eqnarray}

Finally, we note that by using the relation between the incomplete gamma function $\Gamma(a, z)$ and the confluent hypergeometric Whittaker function $U(a, b, z)$ \cite{DLMF}, one can show that the Fourier transform of the function~(\ref{GammaFlatTop}) indeed yields the result~(\ref{FourierFlatTopKummer}) for arbitrary $N$, not necessarily integer.

\subsection{Beam profile in a polar coordinate system}
Above, we considered the profile of a flat-top beam constructed from low-order Hermite--Gaussian modes and therefore having the shape of a smoothed rectangle. For some applications, a radially symmetric flat-top beam may be required. To determine its profile, it is convenient to switch to a polar coordinate system. In polar coordinates $(r, \varphi)$, the eigenmodes of the paraxial wave equation are the Laguerre--Gaussian modes $\LG_{n}^{(l)}$ with radial index $n \ge 0$ and azimuthal index $l \in \mathbb Z$:
\begin{gather}
	\LG_{n}^{(l)}(r, \varphi) = r^{|l|} \exp(-r^2) L_n^{(|l|)}(2r^2) \exp(-il\varphi), \\
	L_n^{(\alpha)}(r) = \frac{r^{-\alpha} \exp(r)}{n!} \frac{d^n\exp(-r) r^{n+\alpha}}{dr^n},\ \alpha \in \mathbb R,
\end{gather}
where $L_{n}^{(\alpha)}(r)$ denotes the associated Laguerre polynomial. We are interested in the radially symmetric case; therefore, we set $l = 0$ and introduce the notation $\LG_n^{(0)}(r, \varphi) = \LG_n(r)$ and $L_n^{(0)} = L_n$. \par
By analogy with Sec.~\ref{sec:theory}, we seek the flat-beam profile $E(r)$ as a superposition of low-order Laguerre--Gaussian modes $\LG_n$ and impose the condition that the first derivatives with respect to $r$ vanish at $r = 0$ (see Eqs.~(\ref{HGsum}--\ref{Normalization})). Since the Laguerre polynomials $L_n$ form a sequence of polynomials whose degree increases by one with $n$, one may transform a sum of Laguerre polynomials with unknown coefficients into an equivalent set of polynomials, in the same manner as the transition from (\ref{SuperposHermite}) to (\ref{SuperposPoly}). The resulting form of the profile $E(r)$ coincides with (\ref{SuperposPoly}) up to the relabeling of the variable $x$ as $r$. The system of equations enforcing vanishing derivatives is likewise unchanged, and its solution therefore follows immediately (see (\ref{HermiteFlatTop}) and (\ref{GammaFlatTop})):
\begin{equation}
	E(r) = \exp(-r^2) \sum_{n=0}^{N/2} \frac{r^{2n}}{n!} = Q(N/2 + 1, r^2). \label{LaguerreFlatTop}
\end{equation}

By analogy with~(\ref{HermiteFlatTopCoeffs}), we obtain an explicit expansion of $E(r)$ in Laguerre--Gaussian modes. To this end, we express $r^{2n}$ in terms of the Laguerre polynomial basis $L_n(r)$~\cite{DLMF}:
\begin{equation}
	r^{2n} = \frac{(n!)^2}{2^n} \sum_{j=0}^n \frac{(-1)^j}{j!(n-j)!} L_j(2 r^2).
\end{equation}
Thus,
\begin{eqnarray}
	E(r) = \exp(-r^2) \sum_{k=0}^{N/2} &&\left( \sum_{n=k}^{N/2} \frac{(-1)^k n!}{2^n k! (n-k)!} \right) \times \nonumber\\ 
    &&\times L_k(2r^2).
\end{eqnarray}

The two-dimensional Fourier transform $\mathcal{F}$ for radially symmetric functions reduces to the zeroth-order \emph{Hankel} transform $\mathcal{H}$:
\begin{equation}
	\mathcal{H}[E(r)](k) = \int_0^\infty E(r) J_0(k r) r\,dr, \label{HankelTransform}
\end{equation}
where $J_0$ is the zeroth-order Bessel function of the first kind. Indeed,
\begin{widetext}
\begin{multline}
	\mathcal{F}[E(r)](k_x, k_y) = \frac{1}{2\pi} \iint_{-\infty}^{\infty} E(\sqrt{x^2+y^2}) \exp(i x k_x + i y k_y) \,dx dy =\\
	= \frac{1}{2\pi} \int_0^\infty dr\,r E(r) \int_{0}^{2\pi} \exp(i k r \cos \varphi) \, d\varphi = \int_0^\infty E(r) J_0(k r) r\,dr,
\end{multline}
\end{widetext}
where, in the final equality, we have used a known integral representation of the Bessel function~\cite{SpecFuncBook_1979} and introduced the notation $k = \sqrt{k_x^2 + k_y^2}$. \par
To compute the Hankel transform of~(\ref{LaguerreFlatTop}), we represent $Q(N/2+1, r^2)$ in terms of the Whittaker function $U$~\cite{DLMF}:
\begin{eqnarray}
	Q(N/2+1, r^2) = \frac{\exp(-r^2)}{\Gamma(N/2+1)} \times \nonumber\\ \times \ U\left(-\frac{N}{2}, -\frac{N}{2}, r^2 \right). \label{QfromU}
\end{eqnarray}
Substituting this expression into the Hankel transform~(\ref{HankelTransform}) and making the substitution $r^2 = t$:
\begin{eqnarray}
	\mathcal H[E(r)](k) = \frac{1}{2\Gamma(N/2+1)} \times&& \nonumber\\ \times \int_0^\infty e^{-t} U\left(-\frac{N}{2}, -\frac{N}{2}, t \right) J_0(k\sqrt{t}) \,dt.&&
\end{eqnarray}
Next, we use the tabulated integral in \cite{DLMF}, which involves the Bessel function:
\begin{eqnarray}
	\mathcal H[E(r)](k) = \frac{\Gamma(N/2+2)\exp(-k^2/4)}{2\Gamma(N/2+1)} \times&& \nonumber\\ \times \ M\left(-\frac{N}{2}, 2, \frac{k^2}{4} \right).&&
\end{eqnarray}
Finally, applying the Kummer transformation~\cite{DLMF} and relabeling $k$ as $r$, we obtain the expression:
\begin{equation}
	\mathcal H[E(r)] = \frac{N+2}{4} M \left( \frac{N+4}{2}, 2, -\frac{r^2}{4} \right).
\end{equation}
For even $N$, the expression simplifies~\cite{DLMF}:
\begin{equation}
	\mathcal H[E(r)] = \frac{1}{2} L_{N/2}^{(1)}\left( \frac{r^2}{4} \right) \exp\left(-\frac{r^2}{4} \right).
\end{equation}
With the normalization $\mathcal H[E(r)](0) = 1$, the obtained expressions should be multiplied by a factor of $4/(N+2)$.
\section{Taylor expansion\label{sec:app:TaylorExpansion}}
Here we present the Taylor-series expansion of the proposed flat-top beam~\eqref{Longitudinal}. In the most general form, the Taylor series of the field $E_{NM}(x,y,z)$ at $x=y=z=0$ reads as follows:
\begin{equation}
E_{NM}(x,y,z)=\sum_{\alpha, \beta, \gamma} \frac{x^\alpha y^\beta z^\gamma}{\alpha!\,\beta!\,\gamma!} \frac{\partial^{\alpha + \beta + \gamma} E_{NM}(0,0,0)}{\partial x^{\alpha} \, \partial y^{\beta} \, \partial z^{\gamma}}.
\end{equation}

The paraxial wave equation derived from the Helmholtz equation under the paraxial approximation,
\begin{equation}
\left( \frac{\partial^2}{\partial x^2} + \frac{\partial^2}{\partial y^2} \right) E_{NM}=4i \frac{\partial E_{NM}}{\partial z},
\end{equation}
provides a convenient expression for the $z$-derivatives:
\begin{equation}
\frac{\partial^{\gamma} E_{NM}}{\partial z^{\gamma}} = \frac{1}{(4\,i)^\gamma} \left( \frac{\partial^2}{\partial x^2} + \frac{\partial^2}{\partial y^2} \right)^{+\!\!\gamma}\!\! E_{NM},
\end{equation}
and for higher mixed derivatives:
\begin{equation}
 \frac{\partial^{\alpha + \beta + \gamma} E_{NM}}{\partial x^{\alpha} \, \partial y^{\beta} \, \partial z^{\gamma}} = \frac{1}{(4\,i)^\gamma}\sum_{j=0}^\gamma \! \begin{pmatrix}\gamma \\ j \end{pmatrix}
\frac{d^{2j+\alpha}E_N}{dx^{2j+\alpha}} \frac{d^{2 \gamma + \beta - 2j}E_M}{dy^{2 \gamma + \beta - 2j}},
\label{HigherCross}
\end{equation}
where we take into account that, by construction, $E_{NM}(x,y,0) = E_N(x) E_M(y)$. This brings us to
\begin{eqnarray}
    &&E_{NM}(x,y,z)=\sum_{\alpha,\beta,\gamma} \sum_{j=0}^\gamma \!
    \begin{pmatrix}\gamma \\ j \end{pmatrix}
    \frac{x^\alpha y^\beta z^\gamma}{\alpha!\,\beta!\,\gamma!}\,\frac{1}{(4\,i)^\gamma} \times \nonumber\\
    &&\times \frac{d^{2j+\alpha}E_N(0)}{dx^{2j+\alpha}} \frac{d^{2 \gamma + \beta - 2j}E_M(0)}{dy^{2 \gamma + \beta - 2j}}.
\label{FTTaylor}
\end{eqnarray}

Let us analyze~\eqref{HigherCross} to find the lowest order $\alpha + \beta + \gamma$ of nonzero terms in~\eqref{FTTaylor}. The derivative~\eqref{HigherCross} is nonzero if both the $x$- and $y$-derivatives are nonzero for some $j$. The orders of nonzero $x$-derivatives are $2j+\alpha = 0, N+2, N+4, \dots$, and those of the nonzero $y$-derivatives are $2\gamma+\beta-2j = 0, M+2, M+4, \dots$. The analysis simplifies when the special cases $\alpha = 0$ and $\beta = 0$ are treated separately. Table~\ref{tab:NonzeroDerivativeOrder} summarizes results. It shows the orders $\alpha + \beta + \gamma$ of nonzero terms for different sets of indices $\alpha$, $\beta$, and $\gamma$.

\begin{table}
    \centering
    \caption{Order $\alpha + \beta + \gamma$ of nonzero derivatives $\displaystyle \frac{\partial^{\alpha+\beta+\gamma} E_{NM}(0,0,0)}{\partial x^{\alpha} \, \partial y^{\beta} \, \partial z^{\gamma}} \ne 0$ in the flat-top Taylor expansion for different sets of indices $\alpha$, $\beta$, and $\gamma$.}
    \label{tab:NonzeroDerivativeOrder}
    \begin{ruledtabular}
    \begin{tabular}{cccl}
    $\alpha$ & $\beta$ & $\gamma$ & $\alpha + \beta + \gamma$ \\
    \hline
    0 & 0 & Nonzero & $\ge \min(N, M)/2 + 1$ \\
    0 & Nonzero & Nonzero &  $\ge M/2 + 2$ \\
    Nonzero & 0 & Nonzero &  $\ge N/2 + 2$ \\
    Nonzero & Nonzero & Nonzero & $\ge (N+M)/2 + 4$ \\
    \end{tabular}
    \end{ruledtabular}
\end{table}

For example, from Table~\ref{tab:NonzeroDerivativeOrder} we find the first terms of the Taylor series for the case $N=M$:
\begin{eqnarray}
   E_{NN}(x,y,z)=1&+&a_1z^{\frac{N}{2}+1}+a_2z^{\frac{N}{2}+2}+\nonumber\\ &+&a_3(x^2+y^2)z^{\frac{N}{2}}+\dots, \label{eq:FlatTopTaylor}
\end{eqnarray}
where the remaining terms are of total order at least $N/2+3$.

We now show how to calculate the coefficients $a_i$. As follows from (\ref{FTTaylor}), it is sufficient to determine the derivatives with respect to $x$ and $y$, after which each corresponding coefficient $a_i$ is obtained by direct summation. Applying power series~\cite[\S8.7.3]{DLMF} of the incomplete gamma function $\Gamma(a,z)$ to Eq.~\eqref{GammaFlatTop}, we find that for even $\alpha \ge N+2$,
\begin{equation}
   \frac{d^{\alpha} E_N(0)}{dx^{\alpha}} = \frac{(-1)^{\alpha/2-N/2} \alpha!} {(N/2)!\, (\alpha/2 - N/2 - 1)! \, \alpha/2}.
\end{equation}
For odd values of $\alpha$ the derivatives are zero. After some algebra, the expression~\eqref{eq:FlatTopTaylor} is rewritten in the form:
\begin{widetext}
\begin{equation}
    E_{NN}(x, y, z) = 1 + \frac{(N+2)! \, z^{N/2}}{(4\,i)^{N/2} (N/2+1)!^2} \left( \frac{i}{2} z - \frac{(N+2)(N+3)}{4(N+4)} z^2 - \frac{N+2}{4} (x^2 + y^2) \right) + \dots.
\end{equation}
\end{widetext}
\section{Parameter measurements \label{sec:app:params}}

In this appendix we explain how the parameters of the numerical model were measured, including the atom temperature, the geometric parameters of the optical traps and the control parameters of the Rydberg lasers. \par
Assuming that the optical trap is formed by a symmetric Gaussian beam, we need to extract the beam-waist radius $w_0$ and the trap depth $U_0$ to fully define the optical trap. We do this by measuring the parametric resonance of the atom survival probability in the dipole trap when its position is periodically modulated. \par
From these measurements we extract radial and axial trap frequencies $\omega_r = 2\pi \times 62~\text{kHz}, \; \omega_z = 2\pi \times 8~\text{kHz}$. This gives us estimates of the trap depth $U_0 = 800~\upmu\text{K}$ and beam-waist radius $w_0=1.4~\upmu\text{m}$. 

\begin{figure}
\includegraphics[width=0.48\textwidth]{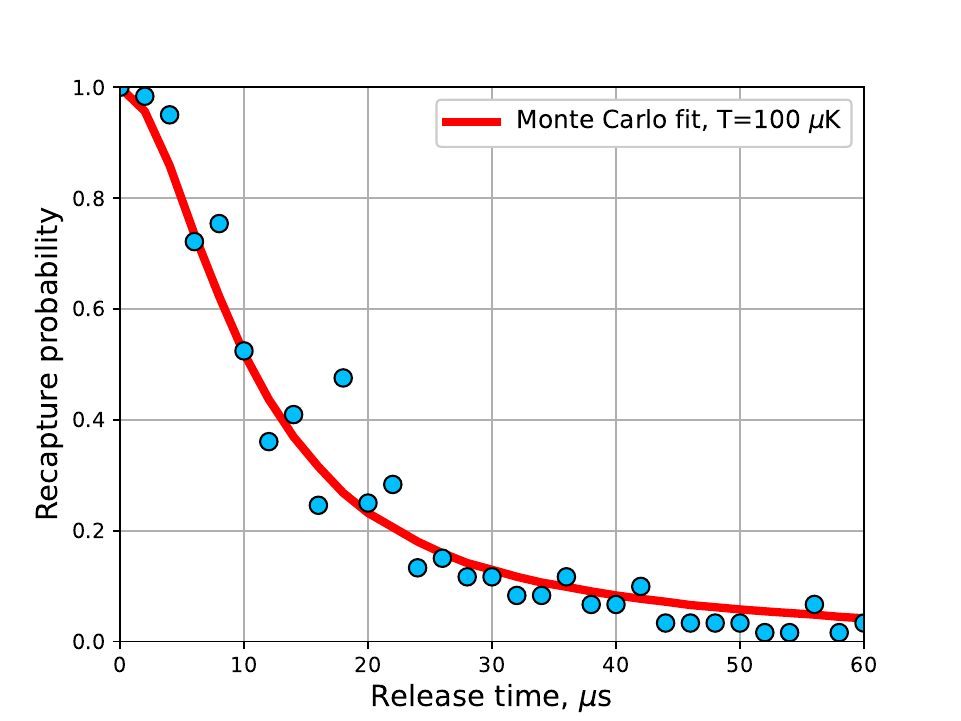}
\caption{\label{Releaserecapture experiment} The atom temperature is extracted by fitting the release--recapture curve with a Monte Carlo simulation.}
\end{figure}

We perform a classic release--recapture  experiment, shown in Fig.~\ref{Releaserecapture experiment} and fit the results with a Monte Carlo simulation to extract the atom temperature $T\approx 100~\upmu\text{K}$.

For modeling Rabi oscillations and CZ-gate implementation, we reconstruct the field profile in the atomic plane from the atomic ionization density map, shown in Figure~\ref{Profiles}a. Assuming that the ionization probability is proportional to the intensity of the blue laser beam, the field profile can be obtained in the following way:
\begin{equation}
   \tilde E(x,y) = E_0 \sqrt{\frac{I(x,y)}{I_0}}, 
\end{equation}
where $I_0$ and $E_0$ are the intensity and the amplitude of field enlighting target atoms, corresponding to the two-photon frequency $\bar \Omega/2\pi = 2.41$ MHz.

In the numerical model, we account for the thermal motion of atoms and consider thr electric field as a sum of Hermite--Gaussian modes defined in \eqref{eq:HGDefinition}:
\begin{equation}
    E(x,y,z) = \sum_{n,m} c_{nm} \mathrm{HG}_{n}(x,z) \mathrm{HG}_{m}(y,z), 
    \label{eq:HGSum}
\end{equation}
where the coefficients $c_{nm}$ are obtained from the decomposition of $\tilde E(x,y)$ assuming that the phase is constant in the plane $z=0$:
\begin{equation}
    c_{nm} = \frac{\iint \tilde E(x,y) \mathrm{HG}_n(x) \mathrm{HG}_m(y)dxdy}{ \left( \iint \mathrm{HG}^2_n(x) \mathrm{HG}^2_m(y) dxdy \right)^{1/2}}.
\end{equation}

The sum~\eqref{eq:HGSum} is infinite and approaches $\tilde E(x,y)$ for any waist $w_0$ chosen to construct the basis of modes in~\eqref{eq:HGDecompositionLongitudinal}. To reduce the computation complexity we consider $n,m < 20$ and select a waist that closely reconstructs the initial field $\tilde E(x,y)$. For the field reconstructed from Fig.~\ref{Profiles}a we use $w_0 = 2\, \upmu$m.

\bibliography{ICQT_flattop}

\end{document}